\documentclass{article}
\usepackage{graphicx} 
\usepackage{geometry}  
\usepackage{amsmath}
\title{Dissipated Correction Map Method with Trapezoidal Rule for the Simulations of Gravitational Waves from Spinning Compact Binary}
\author{Junjie Luo$^{1}$,
Hong-Hao Zhang$^{1}$,
Weipeng Lin,$^{2}$}
\date{August 2024}
\makeatletter 
\begin{document}
\maketitle

DOI:https://doi.org/10.1093/mnras/stae1997

Pubblished by Symmetry\\
$^{1}$ \quad School of Physics, Sun Yat-sen University, Guangzhou 510275, China\\
$^{2}$ \quad School of Physics and Astronomy, Sun Yat-sen University, Zhuhai 519000,  China

\abstract{The correction map method means extended phase-space algorithm with correction map. In our research, we have developed a correction map method, specifically the dissipated correction map method with trapezoidal rule, for numerical simulations of gravitational waves from spinning compact binary systems. This new correction map method, denoted as $CM3$, has shown remarkable performance in various simulation results, such as phase space distance, dissipated energy error, and gravitational waveform, closely resembling the high-order  precision implicit Gaussian algorithm. When compared to the previously used midpoint map which denoted as $C_2$, the $CM3$ consistently exhibits a closer alignment with the highly accurate Gaussian algorithm in waveform evolution and orbital trajectory analysis. Through detailed comparisons and analyses, it is evident that $CM3$ outperforms other algorithms, including $CM2$ and $C_2$ mentioned in this paper, in terms of accuracy and precision in simulating spinning compact binary systems. The incorporation of the trapezoidal rule and the optimization with a scale factor $\gamma$ have significantly enhanced the performance of $CM3$, making it a promising method for future numerical simulations in astrophysics. With the groundbreaking detection of gravitational waves by the LIGO/VIRGO collaboration, interest in this research domain has soared. Our work contributes valuable insights for the application of matched filtering techniques in the analysis of gravitational wave signals, enhancing the precision and reliability of these detection.}

\section{Introduction}
In our recent work, our group has developed correction map for the extended phase-space method for the compact spinning binary systems. Extended phase-space algorithms are proposed by \cite{Pihajoki:2015}, which replicate the origin variables of position and momentum to generate corresponding copy variables, thereby expanding the phase space representation. All of these processes is particularly aimed at enabling the application of second-order explicit leapfrog algorithms to inherently inseparable Hamiltonian systems, exemplified by the complex dynamics of spinning compact binary systems modeled through post-Newtonian approximations as detailed in \cite{Blanchet:2002mb, Tanay2021, Zotos_2019, Li_2020}. By employing such extended phase-space techniques, researchers gain access to numerical solutions not only for the original position and momentum but also for the duplicated variables. In an ideal scenario, the original and replicated variables should maintain strict equality throughout the simulation. However, practical computations reveal that temporal discrepancies tend to accumulate between these paired variables due to their inherent interactions within the system. To counteract this unwanted drift, \cite{Pihajoki:2015} proposes the implementation of a momentum permutation map, which systematically exchanges the values of the original and copied momenta at each time-step, reducing the difference between original variables and copy ones. Therefore, the design of the mapping matrix has a great impact on both the accuracy and stability of the extended phase-space method. Building upon the foundational work laid out in \cite{Pihajoki:2015}, \cite{Liu:2016MNRAS} further refined these developments by devising the coordinate and momenta sequent permutation maps for a fourth-order extended phase-space explicit algorithm. This advanced algorithm was ingeniously crafted using a combination of two Yoshida's triple products derived from the second-order leapfrog algorithm, as initially presented by \cite{Yoshida_1990PRD}. The purpose was to enhance the accuracy in terms of energy error behavior during simulations.

However, despite these improvements, the algorithm exhibited significant shortcomings when applied to numerical simulations of chaotic orbits. Over time, the disparities between the original and extended variables grew due to their mutual interactions, which posed a substantial challenge. While these differences were negligible for non-chaotic or regular orbits, they could be exacerbated in numerically sensitive chaotic systems, leading to a detrimental feedback loop. To address this critical issue, \cite{Luo:2017ApJ} proposed the midpoint map. This method effectively ensures that the original and extended variables remain strictly identical throughout the simulation process. Moreover, it necessitates the use of only a single Yoshida's triple product to construct a fourth-order algorithm, thereby achieving a twofold increase in computational efficiency without compromising the accuracy of results. Moreover, \cite{pan2021extended} successfully employed the midpoint map in the context of coherent post-Newtonian Euler-Lagrange equations, achieving commendable computational outcomes. Further underscoring the efficacy of the midpoint map, \cite{hu2019novel} demonstrated its superior performance relative to various alternative algorithms in their recent study. In order to further improve the accuracy of this algorithm, we design the correction map. The introduction of manifold correction technology into this framework significantly enhances the precision and stability of the algorithms used to simulate such celestial dynamical behaviors. The manifold correction ensures a more accurate preservation of the intrinsic geometric structure of physical systems during long-term evolution, thus reducing cumulative errors that might arise from conventional numerical integration methods and maximizes the energy conservation of the conservative system\cite{Wu_2007APR, Ma:2008ApJ, Wang_2018AAS}. And the correction map mitigates the divergence issues in chaotic orbits.

However, it is important to acknowledge that in reality compact spinning binary stars do indeed experience gravitational dissipation. When gravitational dissipation is taken into account, these systems cease to be conservative and instead continuously emit gravitational waves as they lose energy over time—a prediction rooted in Einstein's General Theory of Relativity. This theoretical prediction was spectacularly confirmed by direct observation in 2016 with the LIGO experiment(\cite{LIGOScientific:2016aoc}), which detected gravitational waves emitted from the coalescence of compact binaries, thereby validating a cornerstone of relativity theory and inaugurating the new era of gravitational wave astronomy. It also arouses the enthusiasm of researchers to study the dynamic evolution of compact celestial bodies. \cite{wu2022explicit} presents such integrators tailored to black hole spacetimes, providing a robust framework for studying the dynamics of particles in these highly curved regions.
\cite{zhou2022note} delve into the specific scenario of charged particles orbiting a Schwarzschild black hole under the influence of an external magnetic field. These studies elucidate the intricate interplay between gravity, charge, and the magnetic field, thereby enriching our understanding of the two-body problem where a charged particle interacts with the black hole.
Ying Wang's work in the papers\cite{wang2021construction} further extends the scope of explicit symplectic integrators to general relativity, focusing on the Hamiltonian associated with Schwarzschild spacetime geometry. The resulting integrators demonstrate their efficacy in accurately simulating the long-term behavior of N-body Hamiltonian systems, as well as in unraveling the chaotic motion of charged particles in the presence of both a black hole and an external magnetic field.
\cite{wu2021construction} addresses the more complex case of Kerr black holes, constructing explicit symplectic integrators within the context of general relativity. To achieve this, a time transformation is ingeniously applied to the Hamiltonian of Kerr geometry, yielding a transformed Hamiltonian composed of five separable components, each with analytically tractable solutions expressed in terms of a new coordinate time. This innovative approach enhances the efficiency and accuracy of numerical simulations for Kerr black hole systems.
Wei Sun's contribution in paper\cite{sun2021applying} introduces an explicit symplectic integrator tailored to the Kerr spacetime geometry, aimed at accurately capturing the non-integrable dynamics of charged particles orbiting a Kerr black hole under the influence of an external magnetic field.

The orbital motion of these compact objects is more dissipative than that of other small objects and the gravitational waves are easier to detect and identify. When the gravitational dissipation term is cosidered, in this thesis we readjust the scheme of the correction map to better serve the numerical simulation of gravitational waveforms. As a comparison, we will introduce an implicit algorithm, there are many kinds of implicit algorithms, for example,  \cite{Suzuki_1990}'s fourth-order composition in this context is realized as a composite of five individual second-order integrators.
Expanding upon this work, \cite{Zhong:2010PRD} incorporated conjugate canonical spin variables (as presented by \cite{Wu_2010APS}) to develop fourth-order canonical explicit-implicit mixed-symplectic methods. In these methods, separable Hamiltonians are handled using the second-order explicit leapfrog algorithm, while nonseparable Hamiltonians are computed using the second-order implicit midpoint method. Subsequent elaborations and applications of this construction were further explored in \cite{Wu_2011} and \cite{Mei_2013EPJC}. \cite{Seyrich:2013PRD} contributed to the field by proposing Gauss Runge-Kutta implicit canonical symplectic schemes, which inherently preserve the underlying structure of the system being modeled. In addition, \cite{Tsang_2015APJL} advanced the development of implicit slimplectic integrators, tailored for the integration of general nonconservative systems, such as a Newtonian two-body problem subject to 2.5PN gravitational radiation reaction terms. These integrators demonstrate the continued evolution and refinement of numerical techniques for accurately simulating complex astrophysical phenomena.

We include the 4-stage Gauss Runge-Kutta implicit canonical symplectic algorithm(\cite{Seyrich:2013PRD}) as a numerical simulation algorithm tool, whose high accuracy numerical solutions can be treated as true solutions. Implicit algorithms are alternative solutions for computing binary star gravitational waves because they do not have the restriction of the use of separable or inseparable Hamiltonian. Nevertheless, implicit algorithms have multiple iterations that consume computational resources heavily, as well as convergence problems of the iterations leading to unavailability of numerical solutions.

The manuscript is structured as follows.
In section~\ref{sec:2}, we present the Hamiltonian formulation for spinning compact binary stars, considering the 2.5PN dissipative term, and present the corresponding gravitational wave equations for the two polarization states. Then the dissipated correction map algorithm with the trapezoidal rule will be introduced, detailing its implementation and underlying principles.
In section~\ref{sec:3}, Initial values for various parameters are specified, and the numerical solutions for spinning compact binary stars are computed using the dissipated correction map algorithm.   The accuracy of this approach is assessed via phase space distance, dissipated energy error, and gravitational waveform comparison, utilizing the numerical solutions from the 4-stage implicit Gaussian algorithm as the reference truth. Finally, conclusions are drawn based on the results and analyses presented throughout the paper in the section~\ref{sec:4}.

\section{Spinning Compact Binary and Gravitational Wave}\label{sec:2}
\subsection{The Hamiltonian of spinning compact binary}

Before introducing the algorithm, we wish to provide a concise description of the Hamiltonian for a spin-aligned compact binary system composed of two objects such as black hole or neutron star, along with the associated formula for gravitational dissipation term. We assume that the individual masses of the two bodies are denoted by \( m_1 \) and \( m_2 \), with their total mass given by \( M = m_1 + m_2 \), the reduced mass defined as \( \mu = \frac{m_1 m_2}{M} \), the mass ratio \( \beta = \frac{m_1}{m_2} \), and the symmetric mass ratio \( \eta = \frac{m_1 m_2}{M^2} = \frac{\beta}{(1+\beta)^2} \). The dimensionless coordinate \( r \) is expressed in units of \( M \), the momentum \( p \) is measured in units of the reduced mass, and \( t \) is denoted as time.

We define the vector \( \mathbf{r} \) to represent the relative separation between the two bodies, with its magnitude representing the distance. The unit vector pointing along \( \mathbf{r} \) is then given by \( \mathbf{n} = \frac{\mathbf{r}}{|\mathbf{r}|} \). 
The provided equations describe the Hamiltonian \( H \) of a binary system, composed of two interacting bodies, taking into account various post-Newtonian (PN) corrections and spin-orbit and spin-spin interactions. The Hamiltonian is expressed as a sum of several sub-Hamiltonians:

\begin{eqnarray} H=H_{N}+H_{1PN}+H_{2PN}+H_{SOSS}. \label{eq:H} \end{eqnarray}

Each of these components represents a distinct aspect of the system's dynamics and is respectively written as:

\begin{eqnarray}
H_{N}=T(\textbf{p})+V(\textbf{r})=\frac{\textbf{p}^{2}}{2}-\frac{1}{r},
\end{eqnarray}

\begin{eqnarray}
H_{1PN} &=& \frac{1}{8}(3\eta-1)\textbf{p}^{4}-\frac{1}{2}[(3+\eta)\textbf{p}^{2} \nonumber \\
&& +\eta(\textbf{n}\cdot\textbf{p})^{2}]\frac{1}{r}+\frac{1}{2r^{2}},
\end{eqnarray}

\begin{eqnarray}
H_{2PN} &=& \frac{1}{16}(1-5\eta+5\eta^{2})\textbf{p}^{6}+\frac{1}{8}[(5-20\eta-
3\eta^{2})\textbf{p}^{4} \nonumber\\ && -2\eta^{2}(\textbf{n}\cdot\textbf{p})^{2}\textbf{p}^{2}-3\eta^{2}
(\textbf{n}\cdot\textbf{p})^{4}]\frac{1}{r} \nonumber\\
&& +\frac{1}{2}[(5+8\eta)\textbf{p}^{2}+3\eta
(\textbf{n}\cdot\textbf{p})^{2}]\frac{1}{r^{2}} \nonumber\\
&& -\frac{1}{4}(1+3\eta)\frac{1}{r^{3}}.
\end{eqnarray}

\( H_{SOSS}=H_{1.5SO}+H_{2SS} \) denote Spin-orbit and spin-spin interactions.
   This term captures the influence of the bodies' intrinsic spins on their orbital motion. It is composed of two parts:
\begin{eqnarray}
H_{1.5SO}=\frac{1}{r^{3}}\textbf{S}\cdot(\textbf{r}\times\textbf{p}),
\end{eqnarray}
\begin{eqnarray}
H_{2SS}=\frac{1}{2r^{3}}[\frac{3}{r^{2}}(\textbf{S}_{0}\cdot\textbf{r})^{2}-\textbf{S}_{0}^{2}],
\end{eqnarray}
where \( \mathbf{S} \) is given by:
\begin{eqnarray}
\mathbf{S} = \left[2 + \frac{3}{2\beta}\right]\mathbf{S}_1 + \left[2 + \frac{3\beta}{2}\right]\mathbf{S}_2,
\end{eqnarray}
and \( \mathbf{S}_0 \) assumes the form:
\begin{eqnarray}
\mathbf{S}_0 = \left(1 + \frac{1}{\beta}\right)\mathbf{S}_1 + \left(1 + \beta\right)\mathbf{S}_2. 
\end{eqnarray}
The positional vector $\mathbf{r}$ and its associated momentum $\mathbf{p}$ obey the Hamiltonian canonical equations, which govern their temporal evolution as follows:
\begin{eqnarray}
\dot{\mathbf{r}} &=& \frac{\partial H}{\partial\mathbf{p}},\\
\dot{\mathbf{p}} &=& -\frac{\partial H}{\partial\mathbf{r}}.
\end{eqnarray}

These equations embody the fundamental dynamics of the non-spin components of the system.Simultaneously, the spin vectors $\mathbf{S}_j$ experience time evolution according to:
\begin{eqnarray}
\dot{\mathbf{S}}_j = \frac{\partial H}{\partial\mathbf{S}_j}\times\mathbf{S}_j, \label{sj}
\end{eqnarray}
wherein the cross product $\times$ encapsulates the intrinsic rotational character of the spin dynamics. Collectively, these expressions define the comprehensive time-dependent behavior of the system as prescribed by the Hamiltonian $H$.

Observing equation \ref{sj}, it becomes evident that neither of the spin variables exhibits the canonical conjugacy property. To rectify this, \cite{Wu_2010APS} capitalized on the conserved nature of the individual spin magnitudes and introduced a novel set of generalized coordinates $\theta_j$ and the corresponding generalized momenta $\xi_j$. This transformation allows for the unit spin vectors to be recast in the following form:
\begin{eqnarray}
\hat{\textbf{S}}_{j}=
\left(\begin{array}{cccc}
\rho_{j}\cos\theta_{j}\\
\rho_{j}\sin\theta_{j}\\
\xi_{j}/S_j
\end{array}\right).
\end{eqnarray}

The Hamiltonian, as expressed by equation \ref{eq:H}, now manifests a structure with ten distinct, canonically conjugate phase space variables, constituted by the tuple $(\mathbf{r}, \theta_1, \theta_2; \mathbf{p}, \xi_1, \xi_2)$. This reconfiguration ensures that each coordinate variable is paired with its proper momentum counterpart, thereby restoring the fundamental symplectic structure inherent in Hamiltonian mechanics. Consequently, under this newfound parametrization, the Hamiltonian assumes the appellation of
\begin{eqnarray}
H(\mathbf{r},\mathbf{\theta_j};
\mathbf{p}, \mathbf{\xi_j})=H(\mathbf{r},
\mathbf{p}, \mathbf{S}_j). \label{eq:nH}
\end{eqnarray}

Before considering gravitational dissipation The constants of motion mentioned in this systems include the total energy $E = H$, total angular moment vector $\mathbf{J} = \mathbf{S}_1 + \mathbf{S}_2 + \mathbf{r} \times \mathbf{p}$ and the spin lengths $\mathbf{S}^2_j = S^2_j$.

These constants of motion play crucial roles in constraining the behavior of the binary system and serve as essential diagnostics when numerically integrating the equations of motion derived from the Hamiltonian (\ref{eq:H}). The sophisticated symplectic integration schemes mentioned above, such as those developed by \cite{Lubich:2010mj}, \cite{Zhong:2010PRD}, \cite{Seyrich:2013PRD}, and \cite{Tsang_2015APJL}, are designed to efficiently and accurately simulate the dynamics of such systems while preserving these fundamental conservation laws.

\subsection{Gravitational radiation reaction}
Before discussing the gravitational dissipation of binary stars, we write down the second-order post-Newtonian approximation Lagrangian corresponding to the Hamiltonian \ref{eq:H}

\begin{eqnarray}
L(\mathbf{r},\dot{\mathbf{r}},\mathbf{S}_1,\mathbf{S}_2)=L_{N}+L_{1PN}+L_{2PN}+L_{SOSS},\label{eq:L}
\end{eqnarray}
where 
\begin{eqnarray}
L_{N}=\frac{\dot{\mathbf{r}}^{2}}{2}+\frac{1}{r}, 
\end{eqnarray}
\begin{eqnarray}
L_{1PN} &=& \frac{1}{8}(1-3\eta)\dot{\mathbf{r}}^{4}+\frac{1}{2}[(3+\eta)\dot{\mathbf{r}}^{2}
\nonumber \\
&& +\eta(\textbf{n}\cdot\dot{\mathbf{r}})^{2}]\frac{1}{r} -\frac{1}{2r^{2}},
\end{eqnarray}

\begin{eqnarray}
L_{2PN}&=&\frac{1}{16}(1-7\eta+13\eta^{2})\dot{\mathbf{r}}^{6}+\frac{1}{8}[(7-12\eta-9\eta^{2})\dot{\mathbf{r}}^{4} \nonumber \\
&&+(4-10\eta)\eta(\textbf{n}\cdot\dot{\mathbf{r}})^{2}\dot{\mathbf{r}}^{2}
+3\eta^{2}(\textbf{n}\cdot\dot{\mathbf{r}})^{4}]\frac{1}{r} \nonumber \\
&&+\frac{1}{2}[(4-2\eta+\eta^{2})\dot{\mathbf{r}}^{2}
+3\eta(1+\eta)(\textbf{n}\cdot\dot{\mathbf{r}})^{2}]\frac{1}{r^{2}} \nonumber \\
&&+\frac{1}{4}(1+3\eta)\frac{1}{r^{3}},
\end{eqnarray}
the terms $L_{SOSS}$ consisted by  $L_{1.5SO}$ and $L_{2SS}$ are provided in \cite{Hartl:2004xr},
\begin{eqnarray}
L_{1.5SO}=-\frac{1}{r^{3}}\textbf{S}\cdot(\textbf{r}\times\dot{\mathbf{r}}),
\end{eqnarray}
\begin{eqnarray}
L_{2SS}=-\frac{1}{2r^{3}}[\frac{3}{r^{2}}(\textbf{S}_{0}\cdot\textbf{r})^{2}-\textbf{S}_{0}^{2}],
\end{eqnarray}
the Langrangian \ref{eq:L} and Hamiltonian \ref{eq:H} can be transformed into each other by Legendre transformation
\begin{eqnarray}
H=\mathbf{p}\cdot \dot{\mathbf{r}}-L,
\end{eqnarray}
\begin{eqnarray}
\mathbf{p}=\partial L/\partial \dot{\mathbf{r}}.
\end{eqnarray}

The Hamiltonian \ref{eq:H} that governs a binary system is conserved, unless dissipative effects from gravitational radiation at the 2.5PN order are taken into account, which render the system non-conservative. Conventionally, such dissipative terms are not directly integrated into the Lagrangian expression of the equation \ref{eq:L}; instead, they should be appended to the equations of motion directly. With doubling the degrees of freedom, $\mathbf{r}\rightarrow (\mathbf{r},\mathbf{\overline{r}})$, \cite{Tsang_2015APJL} introduce the dissipative terms in Lagrangian form:
\begin{eqnarray}
\Lambda(\mathbf{r},\mathbf{\overline{r}},\dot{\mathbf{r}},\dot{\mathbf{\overline{r}}},\mathbf{S}_1,\mathbf{S}_2)
&=& L(\mathbf{r},\dot{\mathbf{r}},\mathbf{S}_1,\mathbf{S}_2)-L(\mathbf{\overline{r}},\dot{\mathbf{\overline{r}}},\mathbf{S}_1,\mathbf{S}_2) \nonumber \\
&&+L_{RR}(\mathbf{r},\mathbf{\overline{r}},\dot{\mathbf{r}},\dot{\mathbf{\overline{r}}}). \label{Lambda}
\end{eqnarray}
Here $\mathbf{r}_{+} = (\mathbf{r} + \mathbf{\overline{r}})/2$ and $\mathbf{r}_{-} = \mathbf{r} - \mathbf{\overline{r}}$. Within the adopted unit system, the description of the radiative reaction force, as presented in the works of \cite{Galley.PhysRevD.79.124027} and \cite{Galley.galley2012radiation}, is given by
\begin{eqnarray}
L_{RR} &=& \eta[\frac{16}{5}\frac{(\dot{\mathbf{r}}_{+}\cdot \textbf{r}_{-})}{|\textbf{r}_{+}|^{4}}-\frac{48}{5}\frac{|\dot{\mathbf{r}}_{+}|^{2} (\dot{\mathbf{r}}_{+}\cdot \textbf{r}_{-})}{|\textbf{r}_{+}|^{3}} \nonumber\\
&& +24\frac{(\dot{\mathbf{r}}_{+}\cdot \textbf{r}_{+})^{2}(\dot{\mathbf{r}}_{+}\cdot \textbf{r}_{-})}{|\textbf{r}_{+}|^{5}}\nonumber\\
&& +\frac{16}{15}\frac{(\dot{\mathbf{r}}_{+}\cdot \textbf{r}_{+})^{2}(\textbf{r}_{+}\cdot \textbf{r}_{-})}{|\textbf{r}_{+}|^{6}}\nonumber\\
&&+\frac{144}{5}\frac{|\dot{\mathbf{r}}_{+}|^{2}(\dot{\mathbf{r}}_{+}\cdot \textbf{r}_{+})^{2}(\textbf{r}_{+}\cdot \textbf{r}_{-})}{|\textbf{r}_{+}|^{5}} \nonumber\\
&& -40\frac{(\dot{\mathbf{r}}_{+}\cdot \textbf{r}_{+})^{3}(\textbf{r}_{+}\cdot \textbf{r}_{-})}{|\textbf{r}_{+}|^{7}}].
\end{eqnarray}
Analogous to the conventional Lagrangian $L$, which yields the conservative Euler-Lagrange equations of motion, the nonconservative Lagrangian $\Lambda$ gives rise to a set of nonconservative Euler-Lagrange equations governing the system's dynamics.
\begin{eqnarray}
[\frac{d}{dt}\frac{\partial\Lambda}{\partial \dot{\mathbf{r}}_{-}}
-\frac{\partial \Lambda}{\partial \mathbf{r}_{-}}]_{PL}=0. \label{eq:dLamda}
\end{eqnarray}

The abbreviation "PL" denotes the physical limit, a condition in which the two variables $\mathbf{r}$ and $\mathbf{\overline{r}}$ are required to be identical (i.e., $\mathbf{r} = \mathbf{\overline{r}}$ and $\dot{\mathbf{r}} = \dot{\mathbf{\overline{r}}}$, or equivalently $\mathbf{r}_{+} = \mathbf{r}$, $\mathbf{r}_{-} = 0$, and $\dot{\mathbf{r}}_{+} = \dot{\mathbf{r}}$) after the necessary derivatives have been computed in equation \ref{eq:dLamda}. It is crucial to note that enforcing the PL in equation \ref{Lambda} would render it devoid of practical utility. When the PL is taken into account, the extended coordinate $\mathbf{\overline{r}}$ and its time derivative $\dot{\mathbf{\overline{r}}}$ cease to feature in the equations.

Although in the PL, $\mathbf{r}$ and $\mathbf{\overline{r}}$ are equated, it is important to emphasize that $\mathbf{\overline{r}}$ is distinct from the extended coordinate $\mathbf{\widetilde{r}}$ which will be introduced in following section and used in the construction of extended phase-space algorithms. Similarly, while $\mathbf{r}$ and $\mathbf{\widetilde{r}}$ are set equal at every integration step, they do not share the same conceptual basis as $\mathbf{r}$ and $\mathbf{\overline{r}}$ in the PL.

Employing the Legendre transformation, \cite{Galley.galley2013classical} derived a New Hamiltonian from the newly formulated non-conservative Lagrangian,
\begin{eqnarray}
& & \Gamma(\mathbf{r},\mathbf{\overline{r}},\mathbf{p},\mathbf{\overline{p}},\mathbf{S}_1,\mathbf{S}_2)
= \mathbf{p}\cdot\dot{\mathbf{r}}-\mathbf{\overline{p}}\cdot\dot{\mathbf{\overline{r}}}
-\Lambda(\mathbf{r},\mathbf{\overline{r}},\dot{\mathbf{r}},\dot{\mathbf{\overline{r}}},\mathbf{S}_1,\mathbf{S}_2) \nonumber \\
&& =H(\mathbf{r},\mathbf{\theta_{j}};
\mathbf{p}, \mathbf{\xi_j})-H(\mathbf{\overline{r}},\mathbf{\theta_{j}};
\mathbf{\overline{p}}, \mathbf{\xi_j}) \nonumber \\
&& ~~ -L_{RR}(\mathbf{r},\mathbf{\overline{r}},\mathbf{p},\mathbf{\overline{p}}). \label{eq:FGamma}
\end{eqnarray}
It is worth mentioning that the momenta $\mathbf{p}$ and $\mathbf{\overline{p}}$ are the standard conjugate momenta associated with the conservative scenario discussed earlier. Furthermore, the term $L_{RR}$ appearing in equation \ref{eq:FGamma} is derived by substituting $\mathbf{p}$ and $\mathbf{\overline{p}}$ for $\dot{\mathbf{r}}$ and $\dot{\mathbf{\overline{r}}}$, respectively, in equation \ref{eq:FGamma}.

The Hamiltonian's canonical equations governing the dynamics of this nonconservative system $\Gamma$ can be expressed as:
\begin{eqnarray}
\dot{\mathbf{r}} &=& \frac{\partial H}{\partial\mathbf{p}}-[\frac{\partial L_{RR}}{\partial\mathbf{p}_{-}}]_{PL}=\mathbf{Q}(\mathbf{r},\theta_{1}, \theta_{2}; \mathbf{p}, \xi_1, \xi_2), \label{eq:dissipR}\\
\dot{\theta}_{1} &=& \frac{\partial H}{\partial\xi_1}=\Theta_1(\mathbf{r},\theta_{1}, \theta_{2}; \mathbf{p}, \xi_1, \xi_2), \\
\dot{\theta}_{2} &=& \frac{\partial H}{\partial\xi_2}=\Theta_2(\mathbf{r},\theta_{1}, \theta_{2}; \mathbf{p}, \xi_1, \xi_2); \\
\dot{\mathbf{p}} &=& -\frac{\partial H}{\partial\mathbf{q}}+[\frac{\partial L_{RR}}{\partial\mathbf{q}_{-}}]_{PL}=\mathbf{P}(\mathbf{r},\theta_{1}, \theta_{2}; \mathbf{p}, \xi_1, \xi_2), \label{eq:dissipP}\\
\dot{\xi}_{1} &=& -\frac{\partial H}{\partial\theta_1}=\Xi_1(\mathbf{r},\theta_{1}, \theta_{2}; \mathbf{p}, \xi_1, \xi_2), \\
\dot{\xi}_{2} &=& -\frac{\partial H}{\partial\theta_2}=\Xi_2(\mathbf{r},\theta_{1}, \theta_{2}; \mathbf{p}, \xi_1, \xi_2). \label{eq:dissipate}
\end{eqnarray}
Indeed, the dissipative terms are conspicuously absent in equation \ref{eq:dissipR}, given that $L_{RR}$ in equation \ref{eq:dissipate} is explicitly independent of the momentum component $\mathbf{p}_{-}$. Analogously to the conservative scenario, the dissipative effects manifest solely in the acceleration equation \ref{eq:dissipP}.

Upon the emission of gravitational radiation, the total time derivative of the Hamiltonian $ H$ fails to vanish, instead assuming the form:
\begin{eqnarray}
\frac{d H}{dt}=\dot{\textbf{r}}\cdot[\frac{\partial L_{RR}}{\partial \mathbf{r}_{-}}]_{PL}. \label{eq:ed}
\end{eqnarray}

Knowing the Hamiltonian canonical equations under the effect of gravitational dissipation, we can obtain numerical solutions for the evolution of binary orbits over time, and these numerical solutions will become the variables used in the calculation of gravitational waves, of course, the actual observed waveform depends on the direction of the wave source and the observer.
Assuming an observer located in the $xoz$ plane, let $\widehat{\mathbf{p}}$ denote the direction along the intersection line of the orbital plane with the horizon. Define $\widehat{\mathbf{q}} = \widehat{\mathbf{N}} \times \widehat{\mathbf{p}}$, where $\widehat{\mathbf{N}}$ represents the orientation of the observer. The two polarization states of a gravitational wave are given by
\begin{eqnarray}
& h_{+}=\frac{1}{2}\left(\widehat{p}_{i} \widehat{p}_{j}-\widehat{q}_{i} \widehat{q}_{j}\right) \widehat{h}^{i j}, \\
& h_{\times}=\frac{1}{2}\left(\widehat{p}_{i} \widehat{q}_{j}+\widehat{p}_{j} \widehat{q}_{i}\right) \widehat{h}^{i j}.
\end{eqnarray}
The indices $i, j = 1, 2, 3$ correspond to the $x$, $y$, and $z$ components, respectively, with repeated indices denoting Einstein's summation convention. The 2PN approximation of the waveform $h_{ij}$ (\cite{will1996gravitational}) expands as 
\begin{eqnarray}
\boldsymbol{h}^{i j}= & \frac{2 \eta M}{D}\left[\widetilde{\boldsymbol{Q}}^{i j}+\boldsymbol{P}^{0.5} \boldsymbol{Q}^{i j}+\boldsymbol{P} \boldsymbol{Q}^{i j}+\boldsymbol{P} \boldsymbol{Q}_{\mathrm{SO}}^{i j}\right. \nonumber\\
& +\boldsymbol{P}^{1.5} \boldsymbol{Q}^{i j}+\boldsymbol{P}^{1.5} \boldsymbol{Q}_{\mathrm{SO}}^{i j}+\boldsymbol{P}^{2} \boldsymbol{Q}^{i j} \nonumber\\
& \left.+\boldsymbol{P}^{2} \boldsymbol{Q}_{\mathrm{SO}}^{i j}+\boldsymbol{P}^{2} \boldsymbol{Q}_{\mathrm{SS}}^{i j}\right]_{\mathrm{TT}}.
\end{eqnarray}
Here, $D$ represents the distance between the observer and the wave source. The superscripts, such as $\boldsymbol{P}^{1.5}$, indicate the effective post-Newtonian (PN) order, while the subscripts specify the nature of each term. The sub-term of $\boldsymbol{h}^{i j}$ are given by:
\begin{equation*}
\begin{split}
\widetilde{\boldsymbol{Q}}^{i j}= & 2\left(v^{i} v^{j}-\frac{M}{r} n^{i} n^{j}\right), 
\end{split}
\end{equation*}

\begin{equation}
\begin{split}
\boldsymbol{P}^{0.5} \boldsymbol{Q}^{i j}= & \frac{\delta m}{M}\{3(\widehat{\mathbf{N}} \cdot \mathbf{n}) \frac{M}{r}[2 n^{(i} v^{j)}-\dot{R} n^{i} n^{j}] \\
& +(\widehat{\mathbf{N}} \cdot \mathbf{v})\left[\frac{M}{r} n^{i} n^{j}-2 v^{i} v^{j}\right]\},
\end{split}
\end{equation}

\begin{equation}
\begin{split}
& \boldsymbol{P} \boldsymbol{Q}^{i j}=\frac{1}{3}\{( 1 - 3 \eta ) [(\widehat{\mathbf{N}} \cdot \mathbf{n})^{2} \frac{M}{r}[(3 v^{2}-15 \dot{R}^{2}+7 \frac{M}{r}) n^{i} n^{j}+\\
& 30 \dot{R} n^{(i} v^{j)}-14 v^{i} v^{j}]+(\widehat{\mathbf{N}} \cdot \mathbf{n})(\widehat{\mathbf{N}} \cdot \mathbf{v})\frac{M}{r}[12 \dot{R} n^{i} n^{j}  \\
&-32 n^{(i} v^{j)}]+(\widehat{\mathbf{N}} \cdot \mathbf{v})^{2}[6 v^{i}v^{j}-2 \frac{M}{r} n^{i} n^{j}]]+[3(1-3 \eta) v^{2}
 \\
& -2(2-3 \eta) \frac{M}{r}] v^{i} v^{j} +4 \frac{M}{r} \dot{R}(5+3 \eta) n^{(i} v^{j)}+\\
& \frac{M}{r}[3(1-3 \eta) \dot{R}^{2}-(10+3 \eta) v^{2}+29 \frac{M}{r}] n^{i} n^{j}\}, 
\end{split}
\end{equation}

\begin{equation}
\begin{split}
&\boldsymbol{P}^{1.5} \boldsymbol{Q}^{i j}=\frac{\delta m}{M}(1-2 \eta)\{(\widehat{\mathbf{N}} \cdot \mathbf{n})^{3} \frac{M}{r}[\frac{5}{4}(3 v^{2}-7 \dot{R}^{2}+\\
& 6 \frac{M}{r}) \dot{R}^{2} n^{i} n^{j}-\frac{17}{2} \dot{R} v^{i} v^{j}-\frac{1}{6}(21 v^{2}-105 \dot{R}^{2}+44 \frac{M}{r}) n^{(i} v^{j)}] \\
& +\frac{1}{4}(\widehat{\mathbf{N}} \cdot \mathbf{n})^{2}(\widehat{\mathbf{N}} \cdot \mathbf{v}) \frac{M}{r}[58 v^{i} v^{j}+(45 \dot{R}^{2}-9 v^{2}\\
& -28 \frac{M}{r}) n^{i} n^{j}-108 \dot{\boldsymbol{r}} n^{(i} v^{j)}]+\frac{3}{2}(\widehat{\mathbf{N}} \cdot \mathbf{n})(\widehat{\mathbf{N}} \cdot \mathbf{v})^{2} \\
& \frac{M}{r}[10 n^{(i} v^{j)}-3 \dot{R} n^{i} n^{j}]+\frac{1}{2}(\widehat{\mathbf{N}} \cdot \mathbf{v})^{3}(\frac{M}{r} n^{i} n^{j}\\
& -4 v^{i} v^{j})\}+\frac{1}{12} \frac{\delta m}{M}(\widehat{\mathbf{N}} \cdot \mathbf{n}) \frac{M}{r}\{2 \boldsymbol { n } ^ { ( i } \boldsymbol { v } ^ { j ) } [\dot{R}^{2}(63 \\
& +54 \eta)-\frac{M}{r}(128-36 \eta)+v^{2}(33-18 \eta)]\\
&+n^{i} n^{j} \dot{R}(\dot{R}^{2}(15-90 \eta)-v^{2}(63-54 \eta) \\
& +\frac{M}{r}(242-24 \eta))-\dot{R} v^{i} v^{j}(186+24 \eta)\}+\\
& \frac{\delta m}{M}(\widehat{\mathbf{N}} \cdot \mathbf{v})\{\frac{1}{2} v^{i} v^{j}[\frac{M}{r}(3-8 \eta)-2 v^{2}(1-5 \eta)] \\
& -n^{(i} v^{j)} \frac{M}{r} \dot{R}(7+4 \eta)-n^{i} n^{j} \frac{M}{r}[\frac{3}{4}(1-2 \eta) \dot{R}^{2}\\
& +\frac{1}{3}(26-3 \eta) \frac{M}{r}-\frac{1}{4}(7-2 \eta)]\},
\end{split}
\end{equation}

\begin{equation}
\begin{split}
& \boldsymbol{P}^2 \boldsymbol{Q}^{i j}=\frac{1}{60}(1-5\eta+5\eta^2)\{24(\widehat{\mathbf{N}}\cdot \mathbf{v})^4(5v^{i} v^{j}-\frac{M}{r}n^{i} n^{j})+ \\
&\frac{M}{r}(\widehat{\mathbf{N}} \cdot \mathbf{n})^4[2(175\frac{M}{r}-465\dot{R}^{2}+93v^2)v^{i} v^{j}+30\dot{R}(63\dot{R}^2 \\
& -50\frac{M}{r}-27v^2)n^{(i} v^{j)}+(1155\frac{M}{r}\dot{R}^2-172(\frac{M}{r})^2-945\dot{R}^4\\
& -159\frac{M}{r}v^2 +630\dot{R}^2v^2-45v^4)n^{i} n^{j}]+24\frac{M}{r}(\widehat{\mathbf{N}} \cdot \mathbf{n})^3(\widehat{\mathbf{N}} \cdot \mathbf{v}) \\
& [87\dot{R}v^i v^j+5\dot{R}(14\dot{R}^2-15\frac{M}{r}-6v^2)n^{i} n^{j} \\
& +16(5 \frac{M}{r}-10 \dot{R}^{2}+2 v^{2}) n^{(i} v^{j)}]+288 \frac{M}{r}(\widehat{\mathbf{N}} \cdot \mathbf{n})(\widehat{\mathbf{N}} \cdot \mathbf{v})^{3} \\
& [\dot{R} n^{i} n^{j}-4 n^{(i} v^{j)}]+24 \frac{M}{r}(\widehat{\mathbf{N}} \cdot \mathbf{n})^{2}(\widehat{\mathbf{N}} \cdot \mathbf{v})^{2}[(35 \frac{M}{r}-45 \dot{R}^{2}+ \\
&9 v^{2}) n^{i} n^{j}-76 v^{i} v^{j}+126 \dot{R} n^{(i} v^{j)}]\} +\frac{1}{15}(\widehat{\mathbf{N}} \cdot \mathbf{v})^{2}\{[5(25-\\
&78 \eta+12 \eta^{2}) \frac{M}{r}-(18-65 \eta+45 \eta^{2}) v^{2}+9(1-5 \eta+5 \eta^{2}) \dot{R}^{2}] \frac{M}{r} n^{i} n^{j} \\
& +3[5(1-9 \eta+21 \eta^{2}) v^{2}-2(4-25 \eta+45 \eta^{2}) \frac{M}{r}] v^{i} v^{j}+\\
&18(6-15 \eta-10 \eta^{2}) \frac{M}{r} \dot{R} n^{(i} v^{j)}\} +\frac{1}{15}(\widehat{\mathbf{N}} \cdot \mathbf{n})(\widehat{\mathbf{N}} \cdot \mathbf{v}) \frac{M}{r}\\
& \{[3(36-145 \eta+150 \eta^{2}) v^{2}-5(127-392 \eta+36 \eta^{2}) \frac{M}{r}\\
&-15(2-15 \eta+30 \eta^{2}) \dot{R}^{2}] \dot{R} n^{i} n^{j}+6(98-295 \eta+30 \eta^{2}) \dot{R} v^{i} v^{j}\\
&+2[5(66-221 \eta+96 \eta^{2}) \frac{M}{r} -9(18-45 \eta+40 \eta^{2}) \dot{R}^{2}-(66-265 \eta \\
&+360 \eta^{2}) v^{2}] n^{(i} v^{j)}\}+\frac{1}{60}(\widehat{\mathbf{N}} \cdot \mathbf{n})^{2} \frac{M}{r} \{[3(33-130 \eta+150 \eta^{2}) v^{4}\\
&+105(1-10 \eta+30 \eta^{2}) \dot{R}^{4}+15(181-572 \eta+84 \eta^{2}) \frac{M}{r} \dot{R}^{2} \\
& -(131-770 \eta+930 \eta^{2}) \frac{M}{r} v^{2}-60(9-40 \eta+60 \eta^{2}) v^{2} \dot{R}^{2}\\
&-8(131-390 \eta+30 \eta^{2}) (\frac{M}{r})^{2}]n^{i} n^{j}+4[(12+5 \eta-315 \eta^{2}) v^{2}\\
&-9(39-115 \eta-35 \eta^{2}) \dot{R}^{2}+5(29-104 \eta +84 \eta^{2}) \frac{M}{r}] v^{i} v^{j}\\
& +4[15(18-40 \eta-75 \eta^{2}) \dot{R}^{2}-5(197-640 \eta+180 \eta^{2}) \frac{M}{r} \\
& +3(21-130 \eta+375 \eta^{2}) v^{2}] \dot{R} n^{(i} v^{j)}\}+ \\
&\frac{1}{60}\{[(467+780 \eta-120 \eta^{2}) \frac{M}{r} v^{2} -15(61-96 \eta+48 \eta^{2}) \frac{M}{r} \dot{R}^{2}\\
& -(144-265 \eta-135 \eta^{2}) v^{4}+6(24-95 \eta+75 \eta^{2}) v^{2} \dot{R}^{2}\\
& -2(642+545 \eta)(\frac{M}{r})^{2}-45(1-5 \eta+5 \eta^{2}) \dot{R}^{4}] \frac{M}{r} n^{i} n^{j}\\
&+[4(69+10 \eta-135 \eta^{2}) \frac{M}{r} v^{2}-12(3+60 \eta+25 \eta^{2}) \frac{M}{r} \dot{R}^{2} \\
& +45(1-7 \eta+13 \eta^{2}) v^{4}-10(56+165 \eta-12 \eta^{2})(\frac{M}{r})^{2}] v^{i} v^{j}\\
&+4[2(36-5 \eta-75 \eta^{2}) v^{2} -6(7-15 \eta-15 \eta^{2}) \dot{R}^{2}\\
& +5(35+45 \eta+36 \eta^{2}) \frac{M}{r}] \frac{M}{r} \dot{R} n^{(i} v^{j)}\},
\end{split}
\end{equation}

\begin{equation}
P Q_{\mathrm{SO}}^{i j}=2(\frac{M}{r})^{2}\{\widehat{\mathbf{N}} \times [(\frac{\delta m}{M})\mathbf{\chi}_s+\mathbf{\chi}_a]\}^{(i} n^{j)},
\end{equation}

\begin{equation}
\begin{split}
& P^{1.5} Q_{\mathrm{SO}}^{i j}=4(\frac{M}{r})^{2}\{3(\mathbf{n} \times \mathbf{v})\cdot [(\frac{\delta m}{M})\mathbf{\chi}_a+\mathbf{\chi}_s]n^{i} n^{j} \\
&[\mathbf{v}\times [(2+\eta)\mathbf{\chi}_s+2(\frac{\delta m}{M})\mathbf{\chi}_a]]^{(i} n^{j)}+\\
& 3\dot{R}[\mathbf{n} \times [\mathbf{\chi}_s+(\frac{\delta m}{M})\mathbf{\chi}_a]]^{(i} n^{j)}-2\eta(\mathbf{n} \times \mathbf{\chi}_s)^{(i} v^{j)}+ \\
& +\eta[(\widehat{\mathbf{N}}\cdot \mathbf{n})\mathbf{v}+2(\widehat{\mathbf{N}}\cdot \mathbf{v})\mathbf{n}-3\dot{R}(\widehat{\mathbf{N}}\cdot \mathbf{n})\mathbf{n}]^{(i}(\widehat{\mathbf{N}} \times \mathbf{\chi}_{\mathrm{s}})^{j)} \},
\end{split}
\end{equation}

\begin{equation}
\begin{split}
& P^{2} Q_{\mathrm{SS}}^{i j}=-6\left(\frac{M}{r}\right)^{3} \eta\left\{\left[\left|\mathbf{\chi}_{\mathrm{s}}\right|^{2}-\left|\mathbf{\chi}_{\mathrm{a}}\right|^{2}-5\left(\mathbf{n} \cdot \mathbf{\chi}_{\mathrm{s}}\right)^{2}+5\left(\mathbf{n} \cdot \mathbf{\chi}_{\mathrm{a}}\right)^{2}\right] n^{i} n^{j}\right. \\
& \left.+2\left[\mathbf{\chi}_{\mathrm{s}}\left(\mathbf{n} \cdot \mathbf{\chi}_{\mathrm{s}}\right)-\mathbf{\chi}_{\mathrm{a}}\left(\mathbf{n} \cdot \mathbf{\chi}_{\mathrm{a}}\right)\right]^{(i} n^{j)}\right\}, 
\end{split}
\end{equation}

where $\delta m=m_1-m_2$, $\mathbf{\chi}_s=\frac{1}{2}(\frac{\mathbf{S}_1}{m_1^2}+\frac{\mathbf{S}_2}{m_2^2})$, $\mathbf{\chi}_a=\frac{1}{2}(\frac{\mathbf{S}_1}{m_1^2}-\frac{\mathbf{S}_2}{m_2^2})$,$\mathbf{v}=(v^1, v^2, v^3)=(1+\frac{1}{2}(3\eta -1)p^2-\frac{(3+\eta)}{r})\mathbf{p}-\eta(\mathbf{n} \cdot \mathbf{p})\frac{n}{r}$ and $\dot{R}=\mathbf{n}\cdot \mathbf{v}$. Onece obtained numerical solutions for the time-evolution of \(\mathbf{q}\) and \(\mathbf{p}\) of a spinning compact binary system through numerical methods, we can construct the corresponding gravitational waveforms. In the subsequent chapter, we will devise an optimized correction map in extended phase-space algorithm for this system.

\subsection{Dissipative correction map in extended phase-space method}
Since the Hamiltonian $H$ can't be separated into multiple integrable parts, the symplectic leapfrog method cannot be applied directly to these Hamiltonians, unless they are suitably modified to a splitting form. An effective approach to solving this problem is the extended phase-space method. \cite{Pihajoki:2015} introduced a new pair of conjugate and canonical variables ($\widetilde{\textbf{r}}$, $\widetilde{\textbf{p}}$) from the original variables $(\mathbf{r}, \mathbf{p})$. This doubles the phase-space variables,(\textbf{r}, \textbf{p})$\rightarrow$(\textbf{r}, $\widetilde{\textbf{r}}$, \textbf{p}, $\widetilde{\textbf{p}}$) and constructs a new Hamiltonian $\widetilde{H}(\mathbf{r}, \mathbf{\widetilde{r}}, \mathbf{p}, \mathbf{\widetilde{p}})$ using two identical Hamiltonians $H_1$ and $H_2$:
\begin{eqnarray}
       \widetilde{H}(\textbf{r},\widetilde{\textbf{r}},\textbf{p},
       \widetilde{\textbf{p}})=H_{1}(\textbf{r},
       \widetilde{\textbf{p}})+H_{2}(\widetilde{\textbf{r}},\textbf{p}). \label{eq:eH}
\end{eqnarray}
When it comes to the Hamiltonian \ref{eq:nH}, the formulation \ref{eq:eH} should be rewrited as:
\begin{eqnarray}
       \widetilde{H}(\mathbf{r}, \mathbf{\theta_j}, \widetilde{\textbf{r}}, \mathbf{\widetilde{\theta_j}}; \mathbf{p}, \mathbf{\xi_j}, \widetilde{\textbf{p}}, \mathbf{\widetilde{\xi_j}})
       =H_{1}(\textbf{r},\mathbf{\theta_j}, 
       \widetilde{\textbf{p}},\mathbf{\widetilde{\xi_j}})+H_{2}(\widetilde{\textbf{r}}, \mathbf{\widetilde{\theta_j}},\textbf{p}, \mathbf{\xi_j}).
\end{eqnarray}

Observing that both $(\mathbf{r}, \mathbf{\theta_j}, \textbf{p}, \mathbf{\xi_j})$ and $(\widetilde{\textbf{r}}, \mathbf{\widetilde{\theta_j}}, \widetilde{\textbf{p}}, \mathbf{\widetilde{\xi_j}})$ constitute two pairs of conjugate canonical variables, it is immediately apparent that the newly formed Hamiltonian $\widetilde{H}$ comprises two distinct integrable components.

Given this property, it is reasonable to anticipate that the second-order leapfrog algorithm would be well suited to numerically whole the Hamiltonian $\widetilde{H}$. The implementation of such splitting methods proceeds as follows: When $\textbf{H}_{1}$ and $\textbf{H}_{2}$ represent the operators that facilitate the analytical solution of the individual Hamiltonians $H_{1}$ and $H_{2}$, respectively, and $h$ denotes the chosen time-step, a standard leapfrog algorithm can be expressed as:
\begin{eqnarray}
\mathbf{A}_2(h)=\textbf{H}_{2}(\frac{h}{2})\textbf{H}_{1}(h)\textbf{H}_{2}(\frac{h}{2}).
\end{eqnarray}

It is noteworthy that, given identical initial conditions, the solution pairs $(\textbf{r}, \textbf{p})$ and $(\widetilde{\textbf{r}}, \widetilde{\textbf{p}})$ are anticipated to coincide at each time step. Nevertheless, their trajectories rapidly deviate in subsequent time steps due to the intricate interdependence between the solution $(\textbf{r}, \widetilde{\textbf{p}})$ derived from $H_{1}$ and the solution $(\widetilde{\textbf{r}}, \textbf{p})$ associated with $H_{2}$.
This phenomenon manifests itself as a compensatory relationship, where any increase in the value of $H_{1}$ over half of the entire Hamiltonian $\widetilde{H}$, is accompanied by a commensurate decrease in $H_{2}$, and vice versa. Motivated by the inherent symmetry between $H_{1}$ and $H_{2}$, \cite{Luo:2017ApJ} devise a midpoint map and ensure the equality between $H_{1}$ and $H_{2}$,
\begin{eqnarray}
\textbf{M}_1=\left(\begin{array}{cccccccc}
 \mathbf{\frac{1}{2}},   \mathbf{\frac{1}{2}},  \textbf{0},  \textbf{0},  \textbf{0},  \textbf{0},  \textbf{0},  \textbf{0} \\
 \mathbf{\frac{1}{2}}, \mathbf{\frac{1}{2}},  \textbf{0},  \textbf{0},  \textbf{0},  \textbf{0},  \textbf{0},  \textbf{0} \\
\textbf{0},  \textbf{0}, \mathbf{\frac{1}{2}}, \mathbf{\frac{1}{2}},  \textbf{0},  \textbf{0},  \textbf{0},  \textbf{0} \\
\textbf{0},  \textbf{0},  \mathbf{\frac{1}{2}},\mathbf{\frac{1}{2}},  \textbf{0},  \textbf{0},  \textbf{0},  \textbf{0} \\
\textbf{0},  \textbf{0},  \textbf{0},  \textbf{0},  \frac{\mathbf{1}}{2}, \frac{\mathbf{1}}{2},  \textbf{0},  \textbf{0} \\
\textbf{0},  \textbf{0},  \textbf{0},  \textbf{0},  \frac{\mathbf{1}}{2}, \frac{\mathbf{1}}{2},  \textbf{0},  \textbf{0} \\
\textbf{0},  \textbf{0},  \textbf{0},  \textbf{0},  \textbf{0},  \textbf{0}, \frac{\textbf{1}}{2},  \frac{\textbf{1}}{2} \\
\textbf{0},  \textbf{0},  \textbf{0},  \textbf{0},  \textbf{0},  \textbf{0}, \frac{\textbf{1}}{2}, \frac{\textbf{1}}{2}
\end{array}\right).\label{M1}
\end{eqnarray}
The purpose of map \( M_1 \) is to take the midpoint values between original variables and their corresponding duplicate variables and reassign these midpoints to both the original and duplicate variables, for example, $\mathbf{r}=\widetilde{\mathbf{r}}=(\mathbf{r}+\widetilde{\mathbf{r}})/2$. This operation effectively aligns initially unequal pairs of original and replica variables, ensuring they become identical and thus preventing further divergence during subsequent evolutionary processes. Subsequently, the leapfrog algorithm combined with the midpoint map can be formulated as:
\begin{eqnarray}
       \mathbf{C}_2(h)= \mathbf{A}_2(h)\textbf{M}_1=\textbf{H}_{2}(\frac{h}{2})\textbf{H}_{1}(h)\textbf{H}_{2}(\frac{h}{2})\textbf{M}_1.
\end{eqnarray}
From the $nth$ time step to the subsequent $(n+1)th$ step, the corresponding numerical solutions can be represented as
\begin{eqnarray}
\left(\begin{array}{cccc}
\textbf{r}\\
\widetilde{\textbf{r}}\\
\mathbf{\theta}_j\\
\widetilde{\mathbf{\theta}}_j\\
\textbf{p}\\
\widetilde{\textbf{p}}\\
\mathbf{\xi}_j\\
\widetilde{\mathbf{\xi}}_J
\end{array}\right)_{n+1}
=\mathbf{C}_2
\left(\begin{array}{cccc}
\textbf{r}\\
\widetilde{\textbf{r}}\\
\mathbf{\theta}_j\\
\widetilde{\mathbf{\theta}}_j\\
\textbf{p}\\
\widetilde{\textbf{p}}\\
\mathbf{\xi}_j\\
\widetilde{\mathbf{\xi}}_J
\end{array}\right)_{n}.
\end{eqnarray}

When gravitational dissipation is taken into account, the constants of motion associated with the Hamiltonian are no longer conserved. Consequently, in evaluating the absolute energy error for Algorithm $\mathbf{C}_2$, we cannot employ the conventional approach \(E - E_0\), where \(E\) is the instantaneous energy and \(E_0\) is the initial energy. Instead, we must resort to Equation \ref{eq:ed}, which accounts for the energy dissipation due to gravitational radiation. To proceed, let us first reexpress Equation \ref{eq:ed} in a suitable form for our purposes.
\begin{eqnarray}
\triangle H(n)=h(\dot{\mathbf{r}}_{n}\cdot[\frac{\partial L_{RR}(r_{n},p_{n})}{\partial \mathbf{r}_{-}}]_{PL}). \label{eq:TH}
\end{eqnarray}
Equation \ref{eq:TH} enables us to compute the energy dissipated between the $(n+1)th$ and $nth$ steps. Alternatively, apart from this approach, we can also estimate the energy dissipated within a single time step by subtracting the Hamiltonian at step $n$ from its value at step $n+1$, i.e., $H_{n+1} - H_n$, utilizing the numerical solutions corresponding to these respective steps. Both methods yield an approximation of the energy loss due to dissipation within a given time interval. However, the integral invariant relation such as equation \ref{eq:ed} can be used as a precision check in numerical integrals because it is more precise than $H_{n+1} - H_n$ (\cite{huang1983accuracy, mikkola2002individual}). Therefore, we can set the absolute energy error of the dissipative system as the following formula:
\begin{eqnarray}
       \triangle EC=(H_0-H_t)-\sum_{n=0}^{\frac{t}{h}-1}\triangle H(n). \label{eq:error1}
\end{eqnarray}
Here $H_0$ is the initial Hamiltonian value, and $H_t$ is the Hamiltonian value at time $t$. Given the presence of numerical errors, the absolute energy error $\triangle EC$ does not equal to zero. However, drawing inspiration from the principles of manifold correction \cite{Wu_2007APR, Ma:2008ApJ, Wang_2018AAS, Luo:2020}, we can devise a correction map matrix:
\begin{eqnarray}
\textbf{M}_2=\left(\begin{array}{cccccccc}
 \mathbf{\frac{\alpha}{2}},   \mathbf{\frac{\alpha}{2}},  \textbf{0},  \textbf{0},  \textbf{0},  \textbf{0},  \textbf{0},  \textbf{0} \\
 \mathbf{\frac{\alpha}{2}}, \mathbf{\frac{\alpha}{2}},  \textbf{0},  \textbf{0},  \textbf{0},  \textbf{0},  \textbf{0},  \textbf{0} \\
\textbf{0},  \textbf{0}, \mathbf{\frac{1}{2}}, \mathbf{\frac{1}{2}},  \textbf{0},  \textbf{0},  \textbf{0},  \textbf{0} \\
\textbf{0},  \textbf{0},  \mathbf{\frac{1}{2}},\mathbf{\frac{1}{2}},  \textbf{0},  \textbf{0},  \textbf{0},  \textbf{0} \\
\textbf{0},  \textbf{0},  \textbf{0},  \textbf{0},  \frac{\mathbf{\alpha}}{2}, \frac{\mathbf{\alpha}}{2},  \textbf{0},  \textbf{0} \\
\textbf{0},  \textbf{0},  \textbf{0},  \textbf{0},  \frac{\mathbf{\alpha}}{2}, \frac{\mathbf{\alpha}}{2},  \textbf{0},  \textbf{0} \\
\textbf{0},  \textbf{0},  \textbf{0},  \textbf{0},  \textbf{0},  \textbf{0}, \frac{\textbf{1}}{2},  \frac{\textbf{1}}{2} \\
\textbf{0},  \textbf{0},  \textbf{0},  \textbf{0},  \textbf{0},  \textbf{0}, \frac{\textbf{1}}{2}, \frac{\textbf{1}}{2}
\end{array}\right),\label{M2}
\end{eqnarray}
where the scaling factor $\alpha$ is obtained through the application of the following relation:
\begin{eqnarray}
       H(\frac{\alpha(\textbf{r}+\widetilde{\textbf{r}})}{2}, \frac{\mathbf{\theta_i}+\widetilde{\mathbf{\theta_i}}}{2}, \frac{\alpha(\textbf{p}+\widetilde{\textbf{p}})}{2}, \frac{\mathbf{\xi_i}+\widetilde{\mathbf{\xi_i}}}{2}) =H_0-\sum_{n=0}^{\frac{t}{h}-1}\triangle H(n). \label{eq:alpha}
\end{eqnarray}
\( M_2 \), in addition to ensuring equality between the original and duplicate variables, serves a crucial function in nullifying the energy error, such that \(\triangle EC = 0 \). Although this does not eliminate the inherent energy error stemming from the  dissipated energy \ref{eq:TH} itself, we can harness the differential structure of the algorithm $A_2$ and incorporate the trapezoidal rule to enhance the accuracy of the dissipated energy calculation. With this modification, Equation \ref{eq:TH} can be rewritten as
\begin{eqnarray}
\triangle H(n)_{TR}=\frac{h}{4}\dot{\widetilde{\mathbf{r}}}_n\cdot[\frac{\partial L_{RR}(\textbf{r}_n,\widetilde{\textbf{p}}_n)}{\partial\mathbf{r}_{-}}]_{PL}+ \nonumber\\ 
 \frac{h}{2}\dot{\mathbf{r}}_{n+\frac{h}{2}}\cdot[\frac{\partial L_{RR}(\widetilde{\textbf{r}}_{\frac{h}{2}},\textbf{p}_{\frac{h}{2}})}{\partial \mathbf{r}_{-}}]_{PL}+ \nonumber\\ 
 \frac{h}{4}\dot{\widetilde{\mathbf{r}}}_{n+1}\cdot[\frac{\partial L_{RR}(\textbf{r}_{n+1},\widetilde{\textbf{p}}_{n+1})}{\partial\mathbf{r}_{-}}]_{PL}. \label{eq:TBH}
\end{eqnarray}
The variables \( \dot{\widetilde{\mathbf{r}}}_n \), \( \textbf{r}_n \), \( \widetilde{\textbf{p}}_n \), \( \dot{\mathbf{r}}_{n+\frac{h}{2}} \), \( \widetilde{\textbf{r}}_{\frac{h}{2}} \), \( \textbf{p}_{\frac{h}{2}} \), \( \dot{\widetilde{\mathbf{r}}}_{n+1} \), \( \textbf{r}_{n+1} \), and \( \widetilde{\textbf{p}}_{n+1} \) have already been computed during thestandard leapfrog algorithm $A_2$ and need not be recalculated. Incorporating Equation \ref{eq:TBH}, we will rewrite \( M_2 \), Equation \ref{eq:error1} and \ref{eq:alpha} as follows:
\begin{eqnarray}
\textbf{M}_3=\left(\begin{array}{cccccccc}
 \mathbf{\frac{\gamma}{2}},   \mathbf{\frac{\gamma}{2}},  \textbf{0},  \textbf{0},  \textbf{0},  \textbf{0},  \textbf{0},  \textbf{0} \\
 \mathbf{\frac{\gamma}{2}}, \mathbf{\frac{\gamma}{2}},  \textbf{0},  \textbf{0},  \textbf{0},  \textbf{0},  \textbf{0},  \textbf{0} \\
\textbf{0},  \textbf{0}, \mathbf{\frac{1}{2}}, \mathbf{\frac{1}{2}},  \textbf{0},  \textbf{0},  \textbf{0},  \textbf{0} \\
\textbf{0},  \textbf{0},  \mathbf{\frac{1}{2}},\mathbf{\frac{1}{2}},  \textbf{0},  \textbf{0},  \textbf{0},  \textbf{0} \\
\textbf{0},  \textbf{0},  \textbf{0},  \textbf{0},  \frac{\mathbf{\gamma}}{2}, \frac{\mathbf{\gamma}}{2},  \textbf{0},  \textbf{0} \\
\textbf{0},  \textbf{0},  \textbf{0},  \textbf{0},  \frac{\mathbf{\gamma}}{2}, \frac{\mathbf{\gamma}}{2},  \textbf{0},  \textbf{0} \\
\textbf{0},  \textbf{0},  \textbf{0},  \textbf{0},  \textbf{0},  \textbf{0}, \frac{\textbf{1}}{2},  \frac{\textbf{1}}{2} \\
\textbf{0},  \textbf{0},  \textbf{0},  \textbf{0},  \textbf{0},  \textbf{0}, \frac{\textbf{1}}{2}, \frac{\textbf{1}}{2}
\end{array}\right).
\end{eqnarray}\label{M3}
\begin{eqnarray}
       H(\frac{\gamma(\textbf{r}+\widetilde{\textbf{r}})}{2}, \frac{\mathbf{\theta_i}+\widetilde{\mathbf{\theta_i}}}{2}, \frac{\gamma(\textbf{p}+\widetilde{\textbf{p}})}{2}, \frac{\mathbf{\xi_i}+\widetilde{\mathbf{\xi_i}}}{2}) =H_0-\sum_{n=0}^{\frac{t}{h}-1}\triangle H(n)_{TR}, \label{eq:gamma}
\end{eqnarray}

\begin{eqnarray}
       \triangle EC2=(H_0-H_t)-\sum_{n=0}^{\frac{t}{h}-1}\triangle H(n)_{TR}. \label{eq:error2}
\end{eqnarray}
The adjustment of $M_3$ to the original and copied variables can be written:
\begin{eqnarray}
\left(\begin{array}{cccc}
\textbf{r}\\
\widetilde{\textbf{r}}\\
\mathbf{\theta}_j\\
\widetilde{\mathbf{\theta}}_j\\
\textbf{p}\\
\widetilde{\textbf{p}}\\
\mathbf{\xi}_j\\
\widetilde{\mathbf{\xi}}_J
\end{array}\right)
=
\left(\begin{array}{cccc}
\gamma (\mathbf{r}+\widetilde{\textbf{r}})/2\\
\gamma (\mathbf{r}+\widetilde{\textbf{r}})/2\\
(\mathbf{\theta}_j+\widetilde{\theta}_j)/2\\
(\mathbf{\theta}_j+\widetilde{\theta}_j)/2\\
\gamma (\mathbf{p}+\widetilde{\textbf{p}})/2\\
\gamma (\mathbf{p}+\widetilde{\textbf{p}})/2\\
(\mathbf{\xi}_j+\widetilde{\mathbf{\xi}}_j)/2\\
(\mathbf{\xi}_j+\widetilde{\mathbf{\xi}}_j)/2
\end{array}\right).
\end{eqnarray}

Compared to \( M_2 \), Mapping \( M_3 \) is designed such that the new energy error formula \ref{eq:error2} consistently equals zero. Additionally, employing the trapezoidal rule for calculating dissipated energy typically yields higher accuracy. Furthermore, Mapping \( M_3 \) endeavors to optimize the numerical solution by incorporating a scale factor \( \gamma \). The extended phase-space algorithms respectively combining with Mapping \( M_2 \) and Mapping \( M_3 \) can be formulated as follows:
\begin{eqnarray}
       \mathbf{CM2}(h)= \mathbf{A}_2(h)\textbf{M}_2=\textbf{H}_{2}(\frac{h}{2})\textbf{H}_{1}(h)\textbf{H}_{2}(\frac{h}{2})\textbf{M}_2,
\end{eqnarray}
\begin{eqnarray}
       \mathbf{CM3}(h)= \mathbf{A}_2(h)\textbf{M}_3=\textbf{H}_{2}(\frac{h}{2})\textbf{H}_{1}(h)\textbf{H}_{2}(\frac{h}{2})\textbf{M}_3.
\end{eqnarray}

The newly designed $CM3$ algorithm will henceforth be referred to as dissipated correction map method with trapezoidal rule.
In the following section, we will conduct a comparative analysis of Algorithms $C_2$, $CM2$, and $CM3$ in the numerical simulations of spinning compact binary systems. We will utilize numerical solutions obtained via the 4-stage implicit Gaussian method, serving as the reference 'truth solution', to assess the performance of these algorithms.
\section{Numerical simulations}\label{sec:3}
Our primary focus lies in assessing the performance of the algorithms outlined in Section \ref{sec:2} when applied to controlling numerical errors in post-Newtonian (PN) systems of spinning compact binaries, as modeled by the Hamiltonian formulation given in Equation \ref{eq:H}. This ten-dimensional canonical spin Hamiltonian possesses four integrals of motion: the total energy and the three components of the total angular momentum vector. However, the lack of a fifth integral renders the system nonintegrable, potentially giving rise to chaotic behavior in certain spin Hamiltonians, as demonstrated in \cite{Zhong:2010PRD}, \cite{Mei_2013EPJC, Mei:2013uqa}, \cite{Luo:2020} and \cite{Luo_2022}.

To this end, we consider orbit 1 with the following initial conditions:
$(\beta;\textbf{r},\textbf{p})=(1;26,0,0,0,0.21,0), $$ \chi_{1}=\chi_{2}=1,
\hat{\textbf{S}}_{1}=(\rho_{1}\cos\frac{\pi}{4},\rho_{1}\sin\frac{\pi}{4},-0.983734), \quad \hat{\textbf{S}}_{2}=(\rho_{2}\cos\frac{\pi}{4},\rho_{2}\sin\frac{\pi}{4},-0.983734),
\rho_{1}=\rho_{2}=\sqrt{1-(-0.983734)^{2}}.$
Here, the spin vectors $\mathbf{S}_{j}=S_{j}\mathbf{\hat{S}}_{j}$ (for $j=1,2$) are defined such that $\mathbf{\hat{S}}_{j}$ are unit vectors and the spin magnitudes $S_{j}=\chi_{j}m_{j}^{2}/m^{2}$, with $0\leq\chi_{j}\leq1$ representing dimensionless spin parameters. Proceeding according to the phase-space expansion procedure detailed in Section \ref{sec:2}, we derive the new Hamiltonian $\widetilde{H}$, thereby allowing the use of the algorithms $C_2$, $CM2$ and $CM3$ in the numerical calculations involving $\widetilde{H}$.

For comparative purposes, an 4-stage implicit Gaussian algorithm with eighth order accuracy, denoted $Gauss4$, will also be employed to solve Hamiltonian \ref{eq:H}, serving as a reference solution.

\begin{figure}
\centering
\includegraphics[width=0.4\textwidth]{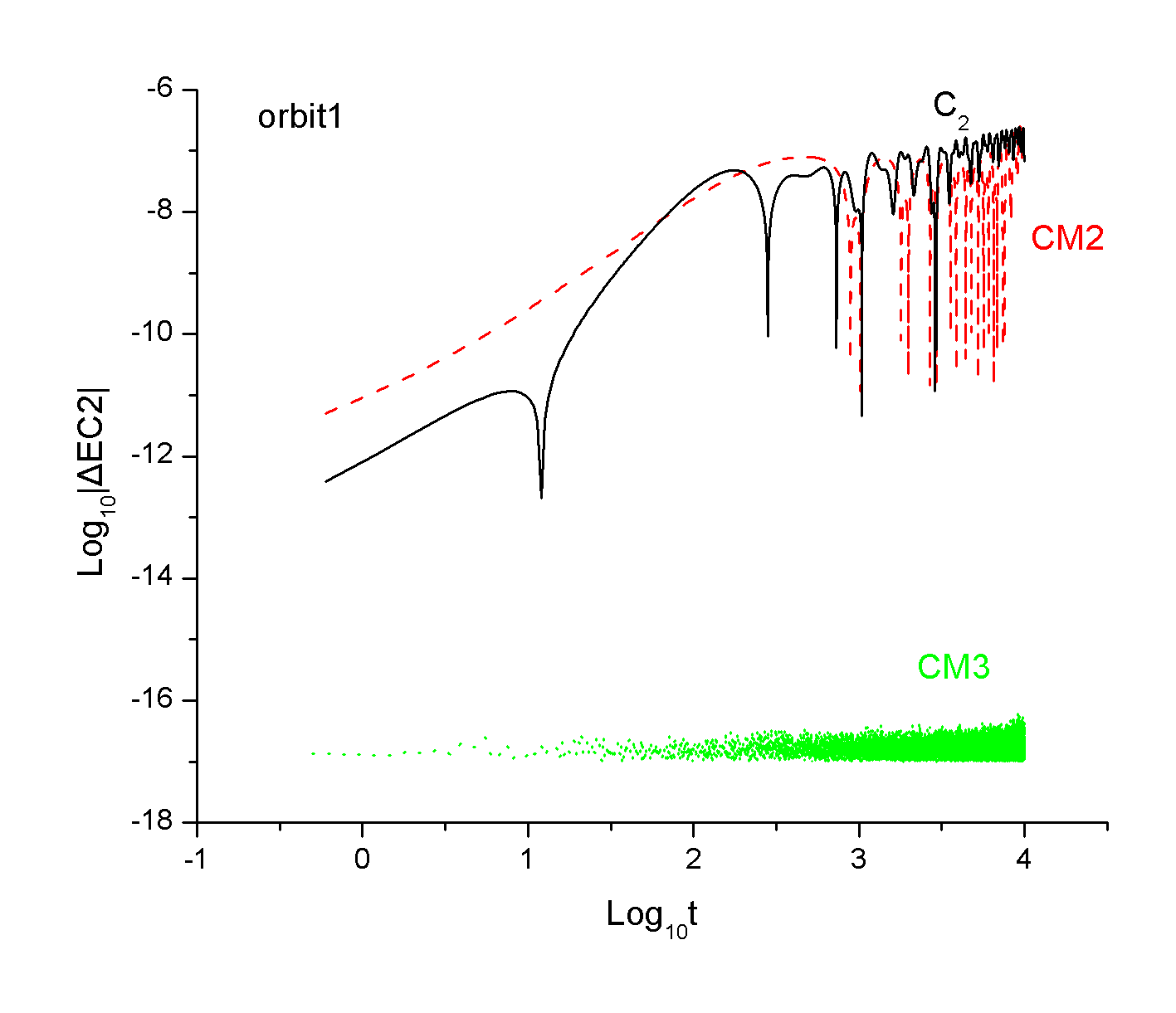}
\caption{Energy error of $\triangle EC2$ calculated by extended phase-space method with all maps. the $CM3$ algorithm (green dot) consistently displays the highest levels of accuracy and long-term stability among the tested methods. Conversely, the energy calculations produced by $C_2$ (black) exhibit the most pronounced bias. The $CM2$ scheme (red dashes) offers a marginal improvement in accuracy compared to $C_2$.}
\label{fig1}
\end{figure}

In Figure \ref{fig1}, we plot the absolute energy errors $\triangle EC2$ for each algorithm with fixed time step $h=0.5$, revealing that $C_2$ exhibits the poorest error behavior, with $CM2$ demonstrating marginally better performance. Both algorithms display a slight energy offset, which is not uncommon in dissipative systems. On the other hand, $CM3$ exhibits the highest precision, approaching the limit of double-precision arithmetic on our computing platform, and has excellent stability. Given $CM3$'s inherent characteristic of correction map $M3$ to minimize $\triangle EC2$, this superior accuracy is anticipated; however, it renders the comparison of energy errors alone insufficient for impartially evaluating the overall performance of algorithms.

\begin{figure}
\centering
\includegraphics[width=0.4\textwidth]{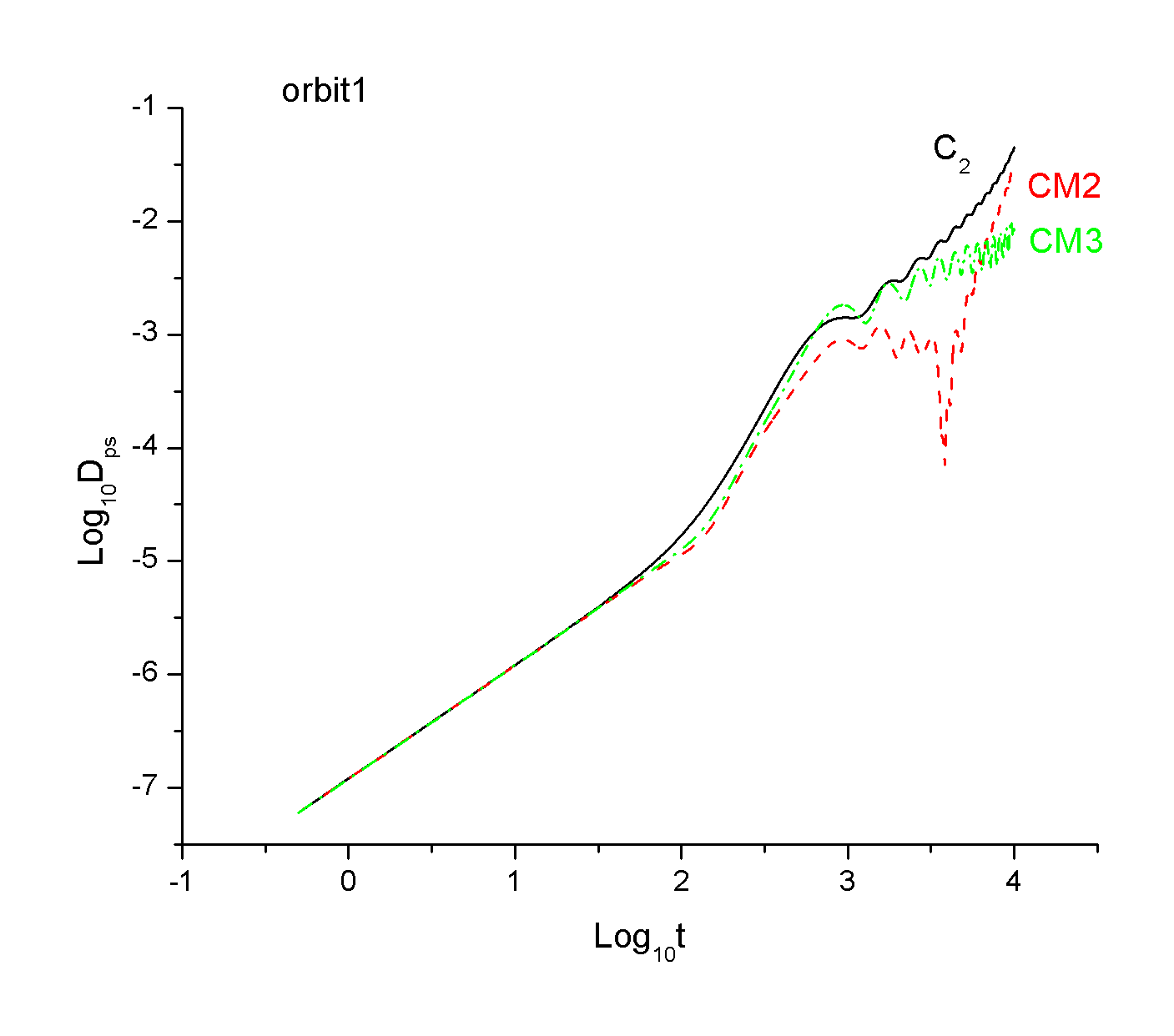}
\caption{Phase space distance $D_{ps}$ between $Gauss4$ and other algorithms as functions of time steps. The distances in the ascending order are $CM3$ (green dot), $CM2$ (red dash), $C_2$ (black). 
The $D_{ps}$ of $CM2$ starts out the smallest distance but keeps growing and eventually becomes the second closest, while $CM3$ rises to the smallest distance. $C_2$ is the longest distance and ranking last.}
\label{fig2}
\end{figure}

To address this, in Figure \ref{fig2}, we present the phase space distances between the numerical solutions produced by each algorithm and the reference solutions obtained using 4-stage implicit Gaussian method. Here $D_{ps}=\sqrt{[(\mathbf{r},\mathbf{\theta_j};
\mathbf{p}, \mathbf{\xi_j})_{gauss}-(\mathbf{r},\mathbf{\theta_j};
\mathbf{p}, \mathbf{\xi_j})]^2}$, the solutions of $Gauss4$ are denoted as $(\mathbf{r},\mathbf{\theta_j};
\mathbf{p}, \mathbf{\xi_j})_{gauss}$. Initially, $CM2$ displays the shortest phase space distance, but it rapidly deteriorates and becomes inferior to $CM3$, which subsequently maintains the closest proximity to the true solution. Throughout this period, $C_2$ consistently remains the farthest from the reference. This plot, while ultimately reaching similar conclusions as Figure \ref{fig1} regarding algorithmic performance, reveals a little differences in the temporal evolution, particularly highlighting $CM2$ as initially outperforming the others.

\begin{figure}
\centering
\includegraphics[width=0.4\textwidth]{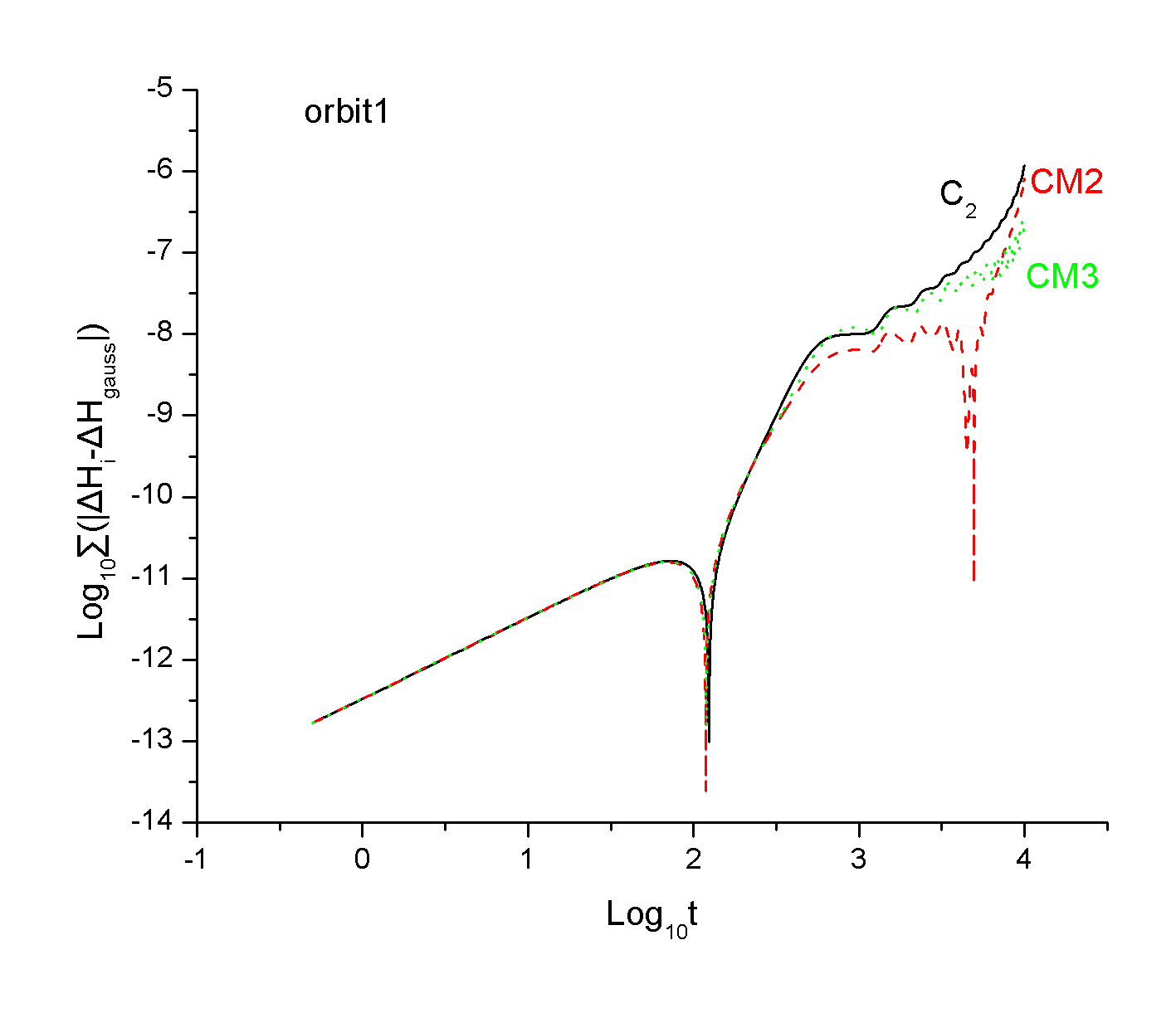}
\caption{Dissipative energy error $\triangle EC3 = \sum_{n=0}^{\frac{t}{h}-1}\triangle H(n)_{TR} - \sum_{n=0}^{\frac{t}{h}-1}\triangle H(n)_{Gauss}$ represents the difference between the dissipated energy$ \sum_{n=0}^{\frac{t}{h}-1}\triangle H(n)_{TR}$ calculated by each algorithm and dissipated energy $\sum_{n=0}^{\frac{t}{h}-1}\triangle H(n)_{Gauss}$ calculated by the higher-order algorithm $Gauss4$. $C_2$ is a solid black line, $CM2$ is represented by a red dash line, and $CM3$ is a green dotted line. It can be seen that the evolution law of each algorithm is highly similar to the phase space distance in Figure $\ref{fig2}$.
}
\label{fig3}
\end{figure}

Beyond phase space distance, we further assess algorithmic performance by subtracting the dissipation energy calculated by each algorithm from that computed using the highly accurate 4-stage implicit Gaussian method, $\triangle EC3 = \sum_{n=0}^{\frac{t}{h}-1}\triangle H(n)_{TR} - \sum_{n=0}^{\frac{t}{h}-1}\triangle H(n)_{Gauss}$. Since the latter provides the most reliable estimate of dissipation energy, this approach offers an objective basis for comparison. In accordance with this strategy, we generate Figure \ref{fig3}. It can be seen that figure \ref{fig3} demonstrates striking similarities in the evolutionary trends with those observed in Figure \ref{fig2}, with $CM2$ initially outperforming its counterparts, followed by a decline in performance and eventual supersession by $CM3$, while $C_2$ consistently underperforms throughout.

\begin{figure}
\centering
\includegraphics[width=0.37\textwidth]{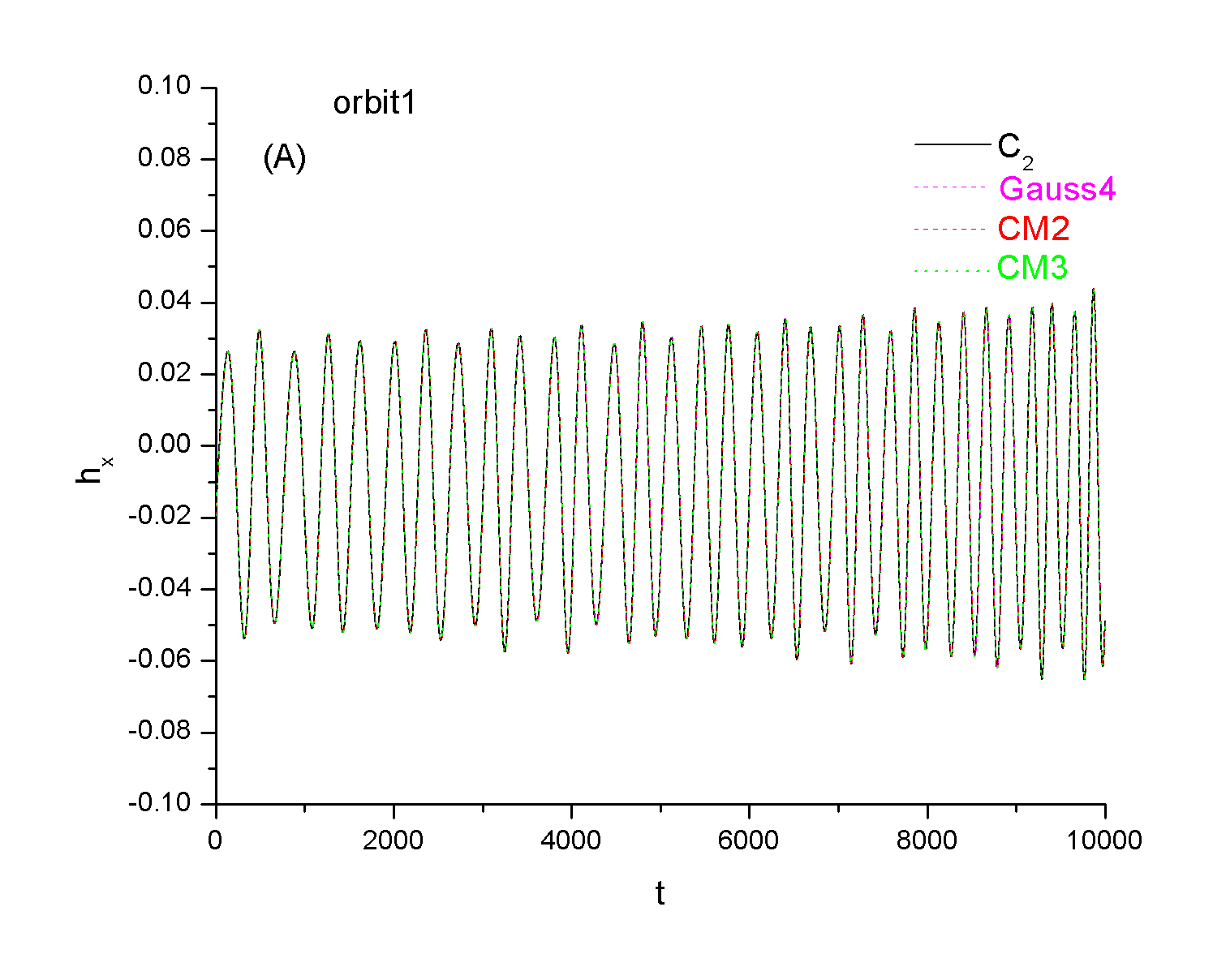}
\includegraphics[width=0.37\textwidth]{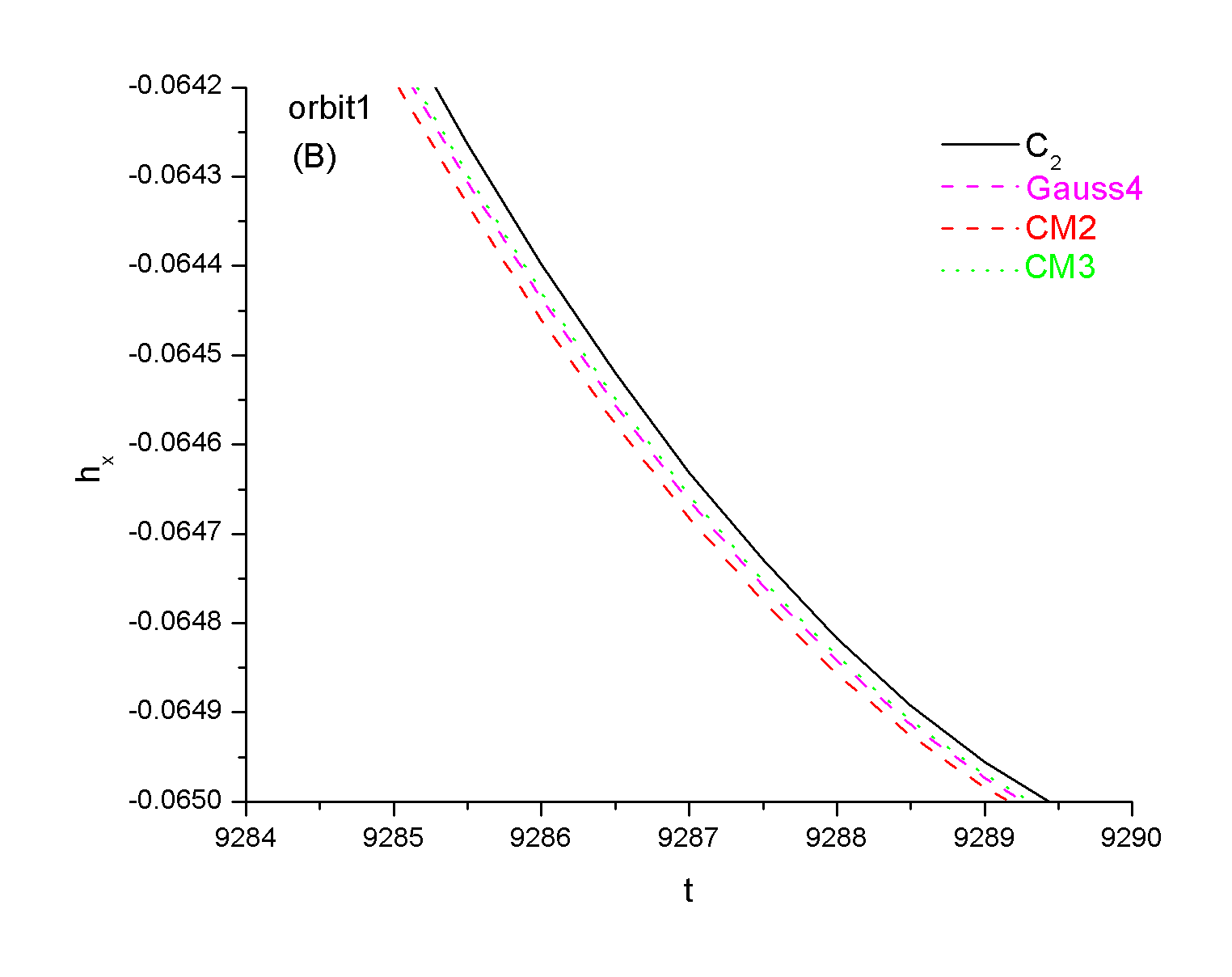}
\includegraphics[width=0.37\textwidth]{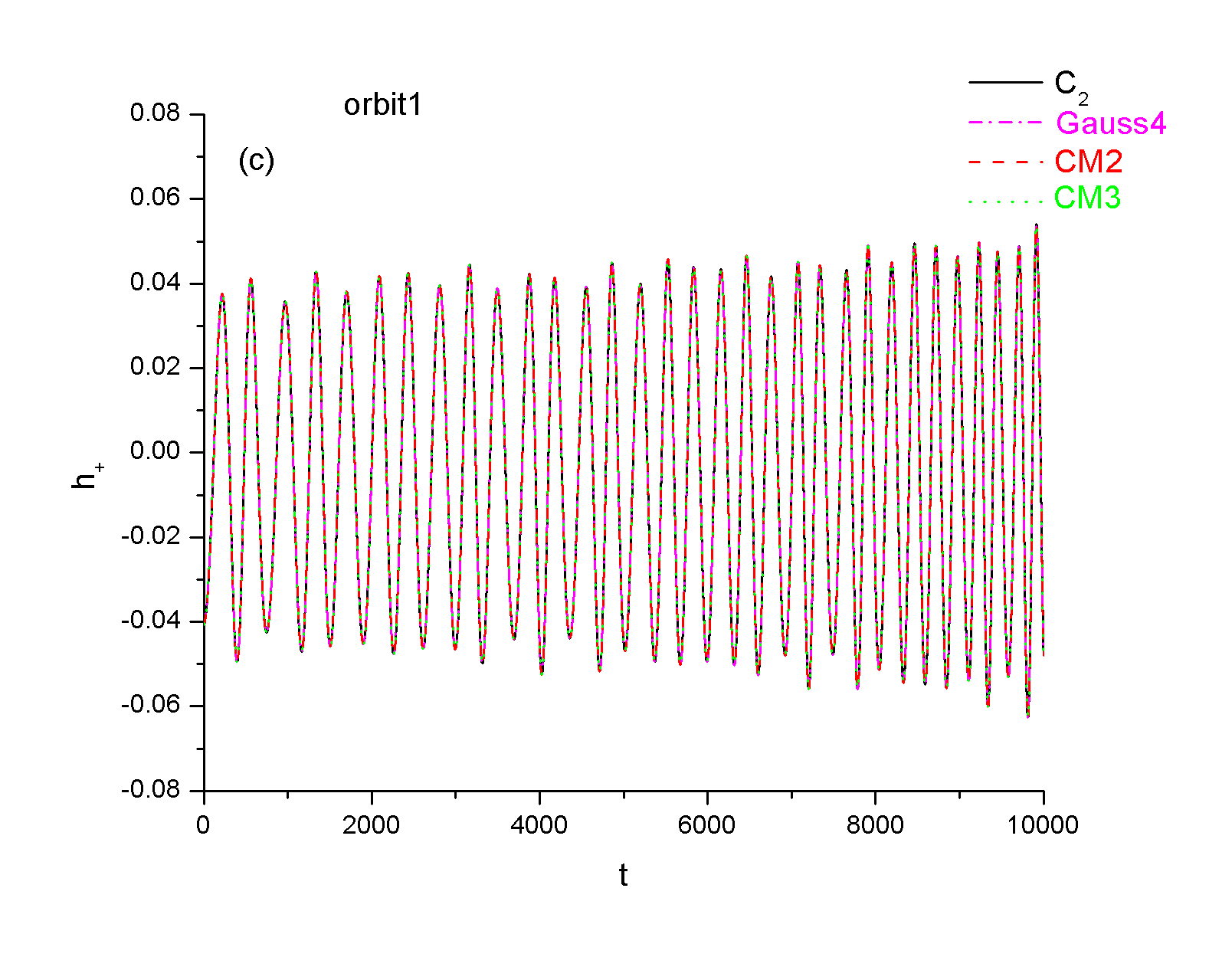}
\includegraphics[width=0.37\textwidth]{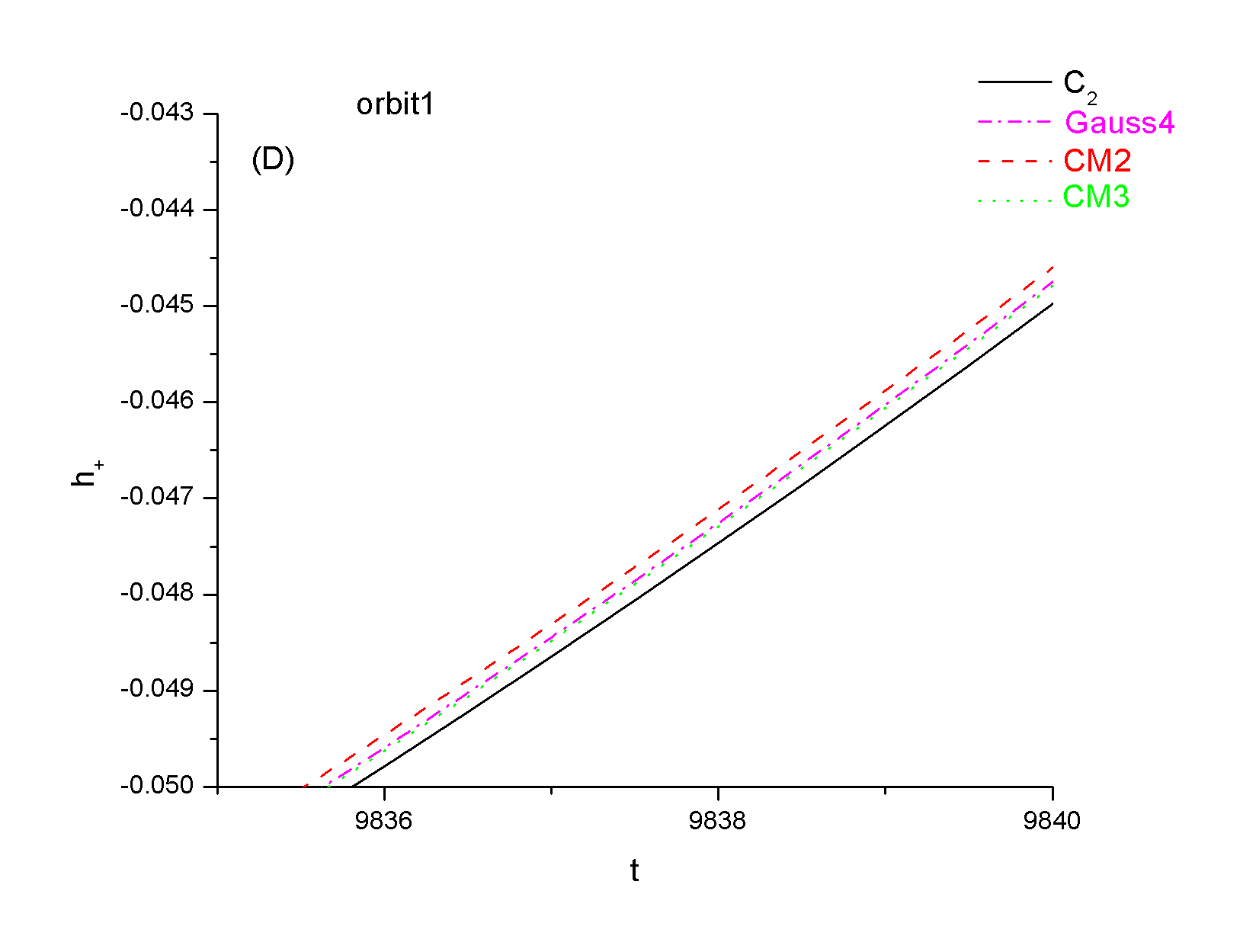}
\caption{The gravitational waveform $h_x$ and $h_+$ in orbit 1. (A). The global evolution diagram of $h_x$ drawn by $C_2$, $CM2$, $CM3$ and $Gauss4$, they are almost the same. (B). $h_x$ local magnification diagram, it can be seen that the result of $CM3$ is closest to $Gauss4$, followed by $CM2$, and finally $C_2$. (C). The global evolution graph of $h_ +$, with almost no difference among all algorithms. (D). $h_+$ locally enlarged figure shows that the rankings of evolution closest to $Gauss4$ are $CM3$, $CM2$, and $C_2$.}
\label{fig4}
\end{figure}

Figure \ref{fig4} presents gravitational waveform plots generated by various algorithms with setting the direction $\widehat{\mathbf{p}} = (1, 0, 0)$ and the orientation of the observer $\widehat{\mathbf{N}} = (0, \sin(\pi/4), \cos(\pi/4))$. Specifically, Figure \ref{fig4}a depicts the waveform for $h_x$, where it is immediately apparent that the waveforms produced by $C2, CM2, CM3$, and $Gauss4$ are virtually indistinguishable to the naked eye due to their near-complete overlap. However, upon closer inspection through local magnification, discernible differences become evident. Figure \ref{fig4}b serves this purpose, providing an amplified view of the $h_x$ waveform evolution. Here, it is clear that $CM3$ yields the waveform closest in resemblance to that produced by the high-order $Gauss4$ algorithm, followed by $CM2$, with $C_2$ exhibiting the greatest deviation from $Gauss4$.

Figure \ref{fig4}c then displays the $h_y$ component of the gravitational waves, generated by each algorithm. Similarly, at a cursory glance, the waveforms appear largely similar. To reveal the nuances, Figure \ref{fig4}d offers a magnified look at the local details of the $h_y$ waveforms, again revealing that $CM3$ maintains the closest alignment with the highly accurate $Gauss4$ waveform, while the other algorithms follow in decreasing proximity.

\begin{figure}
\centering
\includegraphics[width=0.45\textwidth]{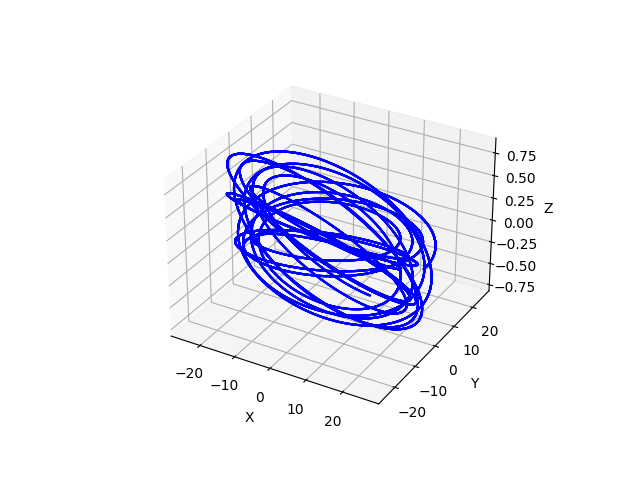}
\caption{The projection of orbit 1 onto $x-y-z$ space plot by $Gauss4$.}
\label{fig5}
\end{figure}

Lastly, Figure \ref{fig5} presents a graphical rendering of the binary star's orbital trajectory in configuration space as calculated by $Gauss4$. This visual representation serves to provide readers with a more intuitive understanding of the underlying physical dynamics, complementing the waveform analysis and offering a comprehensive perspective on the system's behavior as modeled by the highest precision algorithm under consideration.

\begin{figure}
\centering
\includegraphics[width=0.4\textwidth]{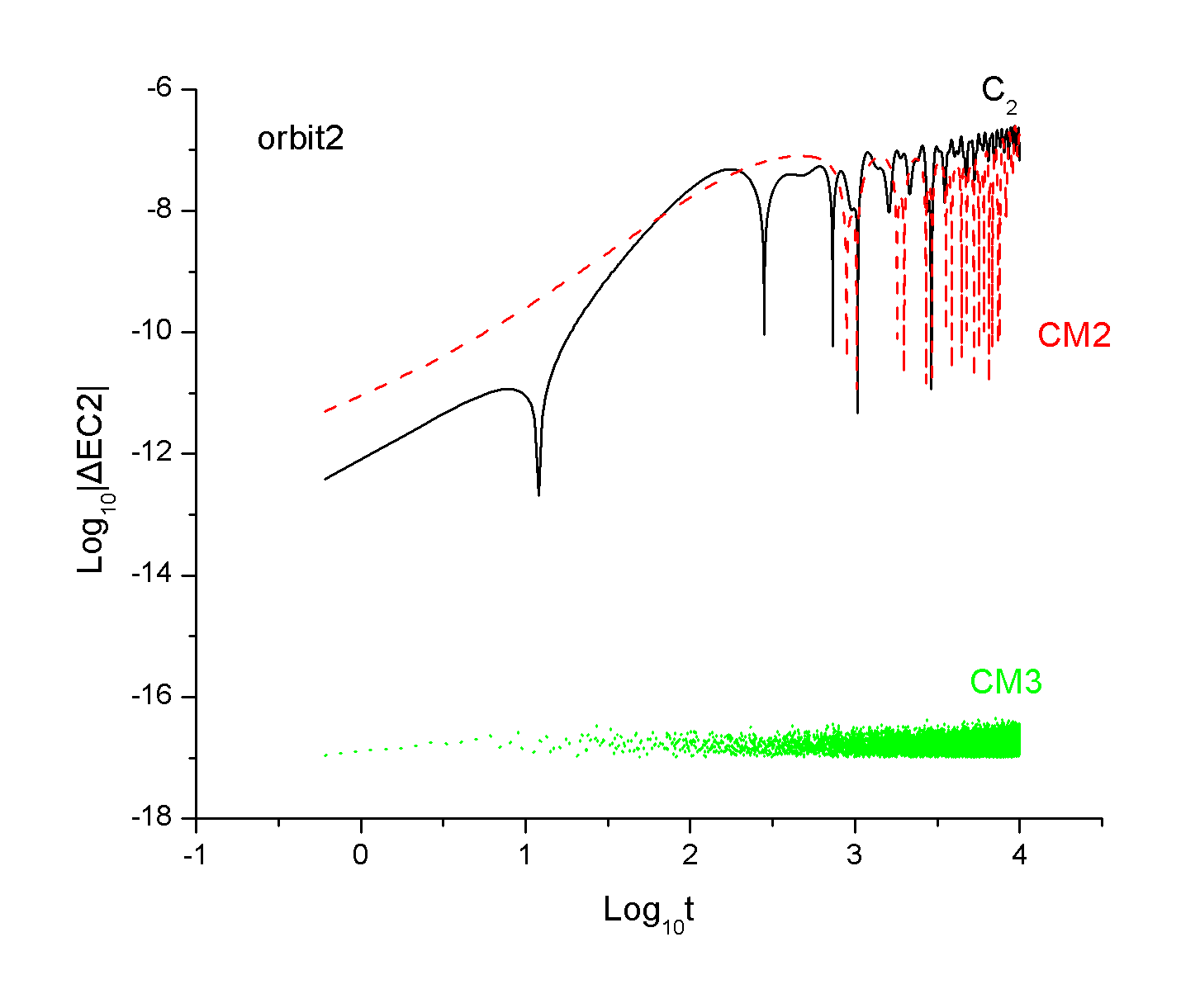}
\caption{The extended phase-space method, incorporating all maps, was utilized to compute the energy error of $\triangle EC2$. In particular, the $CM3$(green dot) algorithm consistently demonstrated the highest degrees of precision and long-term stability among the methods tested. On the other hand, the energy calculations generated by $C_2$ (black) exhibited the most significant deviation. The $CM2$(red dash) scheme offered a slight enhancement in accuracy compared to $C_2$.}
\label{fig6}
\end{figure}

\begin{figure}
\centering
\includegraphics[width=0.4\textwidth]{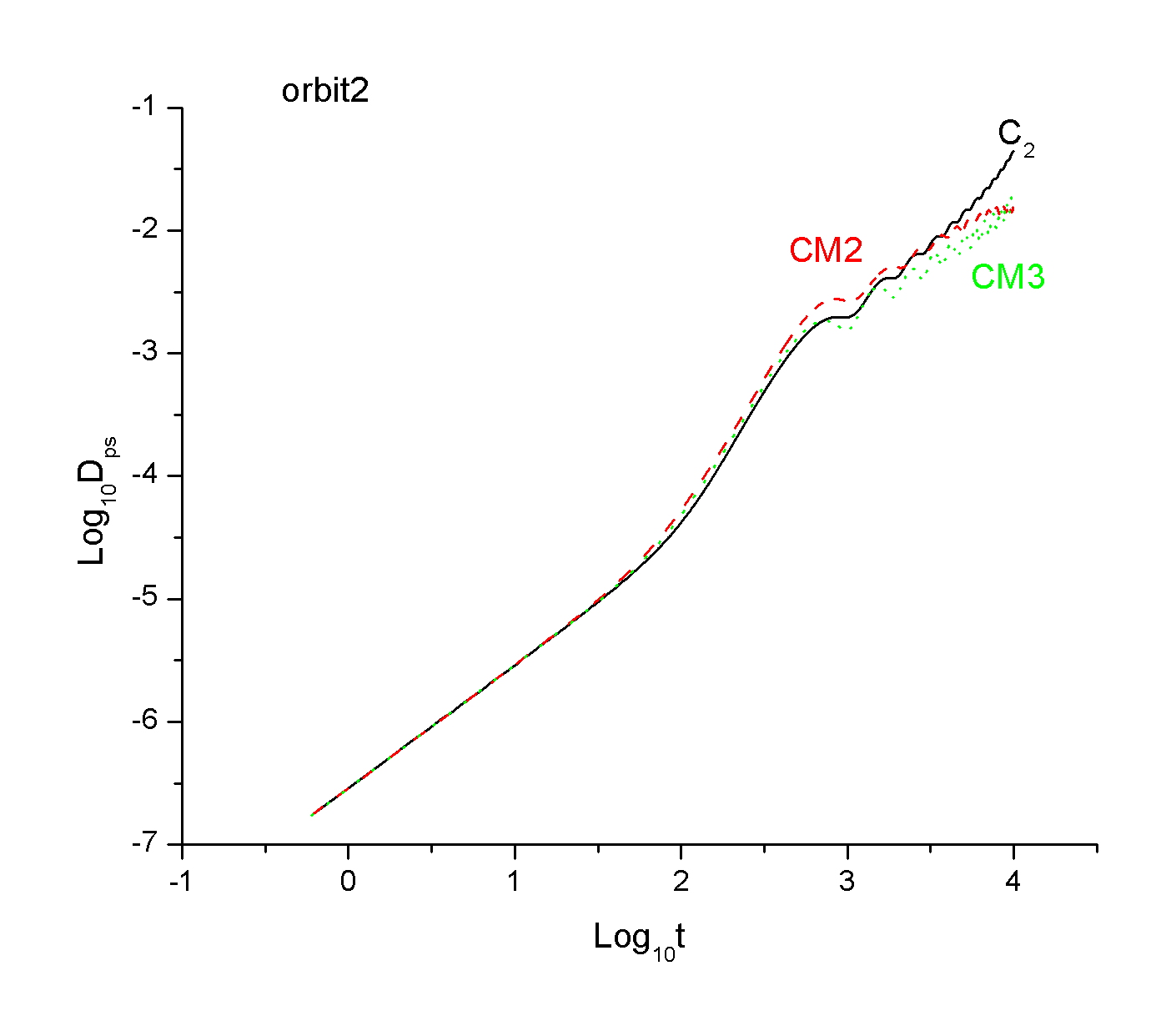}
\caption{Phase space distance $D_{ps}$ between $Gauss4$ and other algorithms as functions of time steps. The distances in the ascending order are $CM3$ (green dot), $CM2$ (red dask), $C_2$ (black). 
The $D_{ps}$ of $CM3$ keeps the smallest distance, while $CM2$ is farther away. $C_2$ is the longest distance and ranking last.}
\label{fig7}
\end{figure}

To delve deeper into the comparative performance of these algorithms, we conduct numerical simulations for a distinct orbit, designated Orbit 2, characterized by the following initial conditions:$(\beta;\textbf{r},\textbf{p})=(1;23,0,0,0,0.24,0), $$ \chi_{1}=\chi_{2}=1,
\hat{\textbf{S}}_{1}=(\rho_{1}\cos\frac{\pi}{4},\rho_{1}\sin\frac{\pi}{4},-0.983734), \quad \hat{\textbf{S}}_{2}=(\rho_{2}\cos\frac{\pi}{4},\rho_{2}\sin\frac{\pi}{4},-0.983734),
\rho_{1}=\rho_{2}=\sqrt{1-(-0.983734)^{2}}.$ and setting fix time-step $h=0.6$
While there is a degree of consistency in the overall performance of each algorithm between Orbit 1 and Orbit 2, subtle differences emerge, except the patterns of energy errors. From figure \ref{fig6} we see that the energy error  $\triangle EC2$ evolution law drawn in orbit 2 and orbit 1 has the same conclusion, and $CM3$ is still the most accurate and stable long-term evolution, $CM2$ ranks second with a gap of several orders of magnitude, and $C_2$ ranks last slightly below $CM2$. The change of the phase space distance diagram in Figure \ref{fig7} compared for orbit 1 is that $CM3$ maintains the minimum distance from the exact solution. $CM2$ does not have an advantage like orbit 1 at the beginning and then lower than $CM3$, but slightly worse than $CM3$ in the long run. Of course, both $CM2$ and $CM3$ outperform $C_2$ in approximating the exact phase space behavior, the comparison of their performances in orbit 2 reveals that $CM3$ sustains its role as the superior integrator. 

\begin{figure}
\centering
\includegraphics[width=0.4\textwidth]{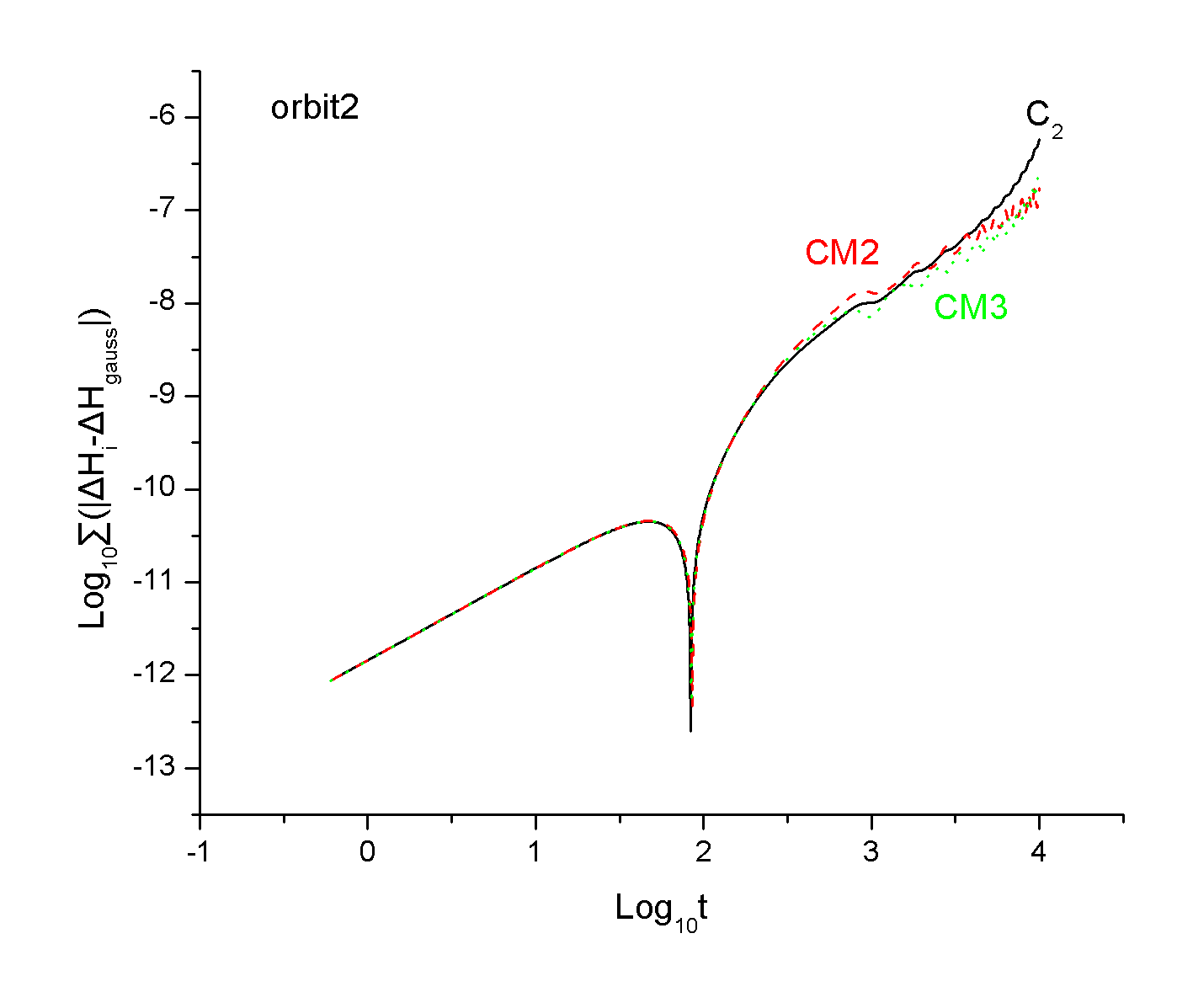}
\caption{Dissipative energy error $\triangle EC3 = \sum_{n=0}^{\frac{t}{h}-1}\triangle H(n)_{TR} - \sum_{n=0}^{\frac{t}{h}-1}\triangle H(n)_{Gauss}$ represents the difference between the dissipated energy$ \sum_{n=0}^{\frac{t}{h}-1}\triangle H(n)_{TR}$ calculated by each algorithm and dissipated energy $\sum_{n=0}^{\frac{t}{h}-1}\triangle H(n)_{Gauss}$ calculated by the higher-order algorithm $Gauss4$. The evolution of $C_2$ (black), $CM2$ (red dash) and $CM3$ (green dot) are highly similar to the phase space distance in Figure $\ref{fig7}$.
}
\label{fig8}
\end{figure}

Consistent with the observations from the phase space distance analysis, the energy error diagram in Figure \ref{fig8}, which employs the $Gauss4$-calculated dissipative energy as the reference truth value, further confirms the performance hierarchy of the investigated algorithms. Once again, $CM3$ emerges as the most accurate, followed closely by $CM2$, while $C_2$ occupies the third position. The evolutionary trends displayed in this energy error plot strikingly resemble those encountered in Figure \ref{fig7}, highlighting the strong correlation between the phase space distance and energy error metrics in characterizing the efficacy of these numerical integration schemes, whether it's orbit 1 or orbit 2. The congruity between the phase space distance and energy error profiles underscores the fact that both measures are effectively capturing the same fundamental aspect of algorithmic performance: the ability to accurately track the true dynamics of the spinning compact binary system. This strong correlation implies that the conclusions drawn from analyzing any one indicator in isolation are likely to echo those of the other, enhancing the reliability of the overall assessment.

\begin{figure}
\centering
\includegraphics[width=0.4\textwidth]{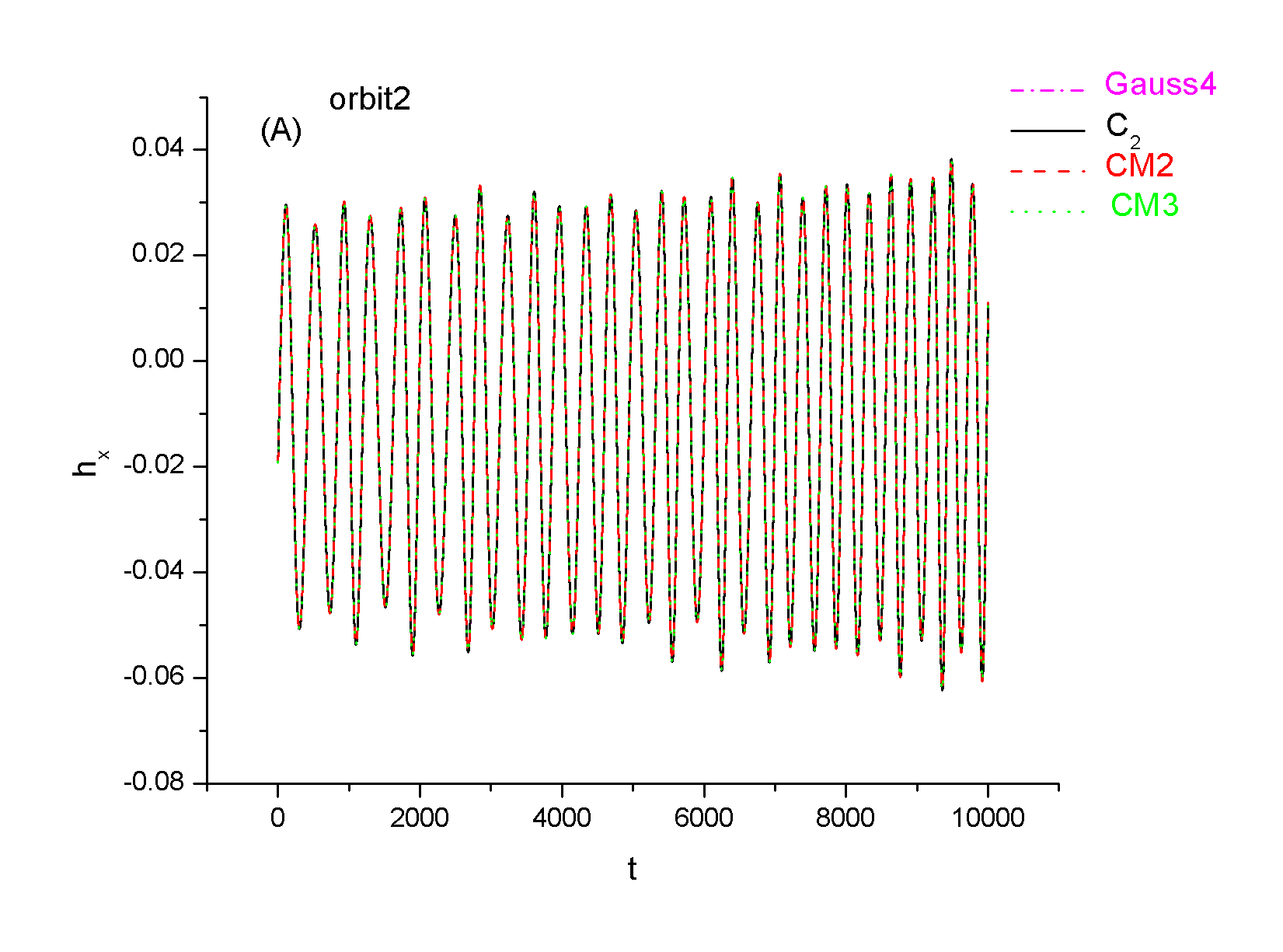}
\includegraphics[width=0.4\textwidth]{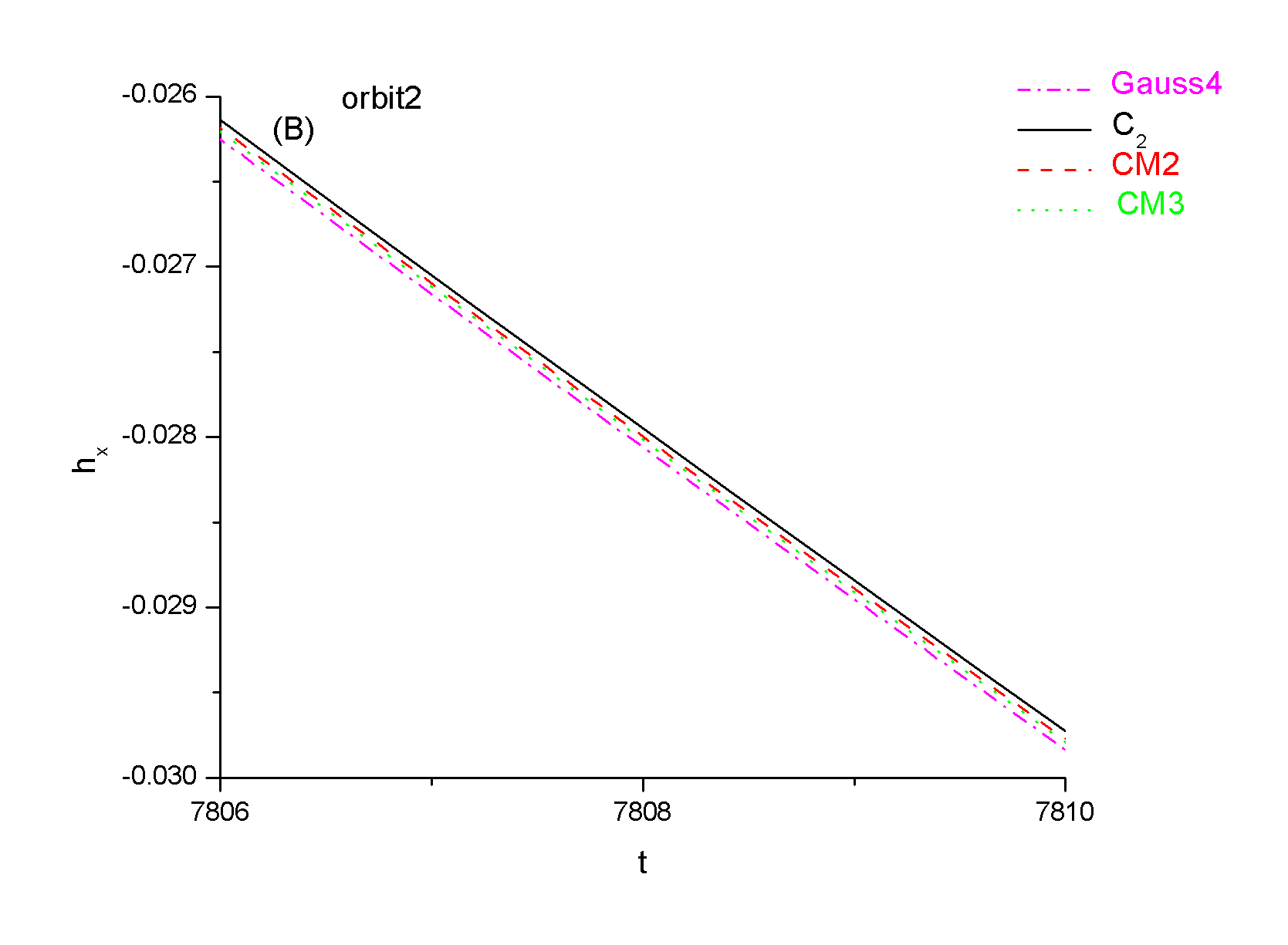}
\includegraphics[width=0.4\textwidth]{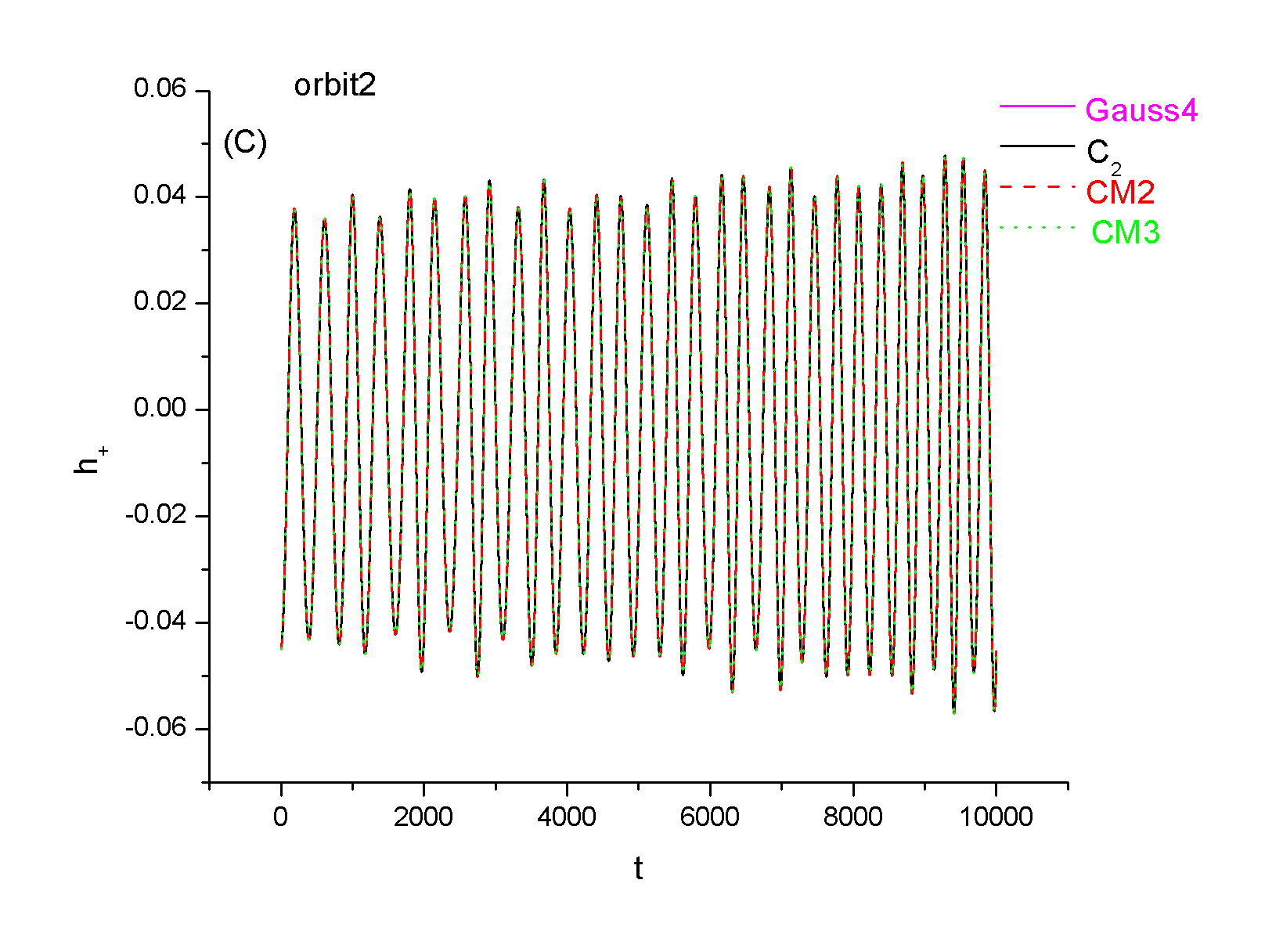}
\includegraphics[width=0.4\textwidth]{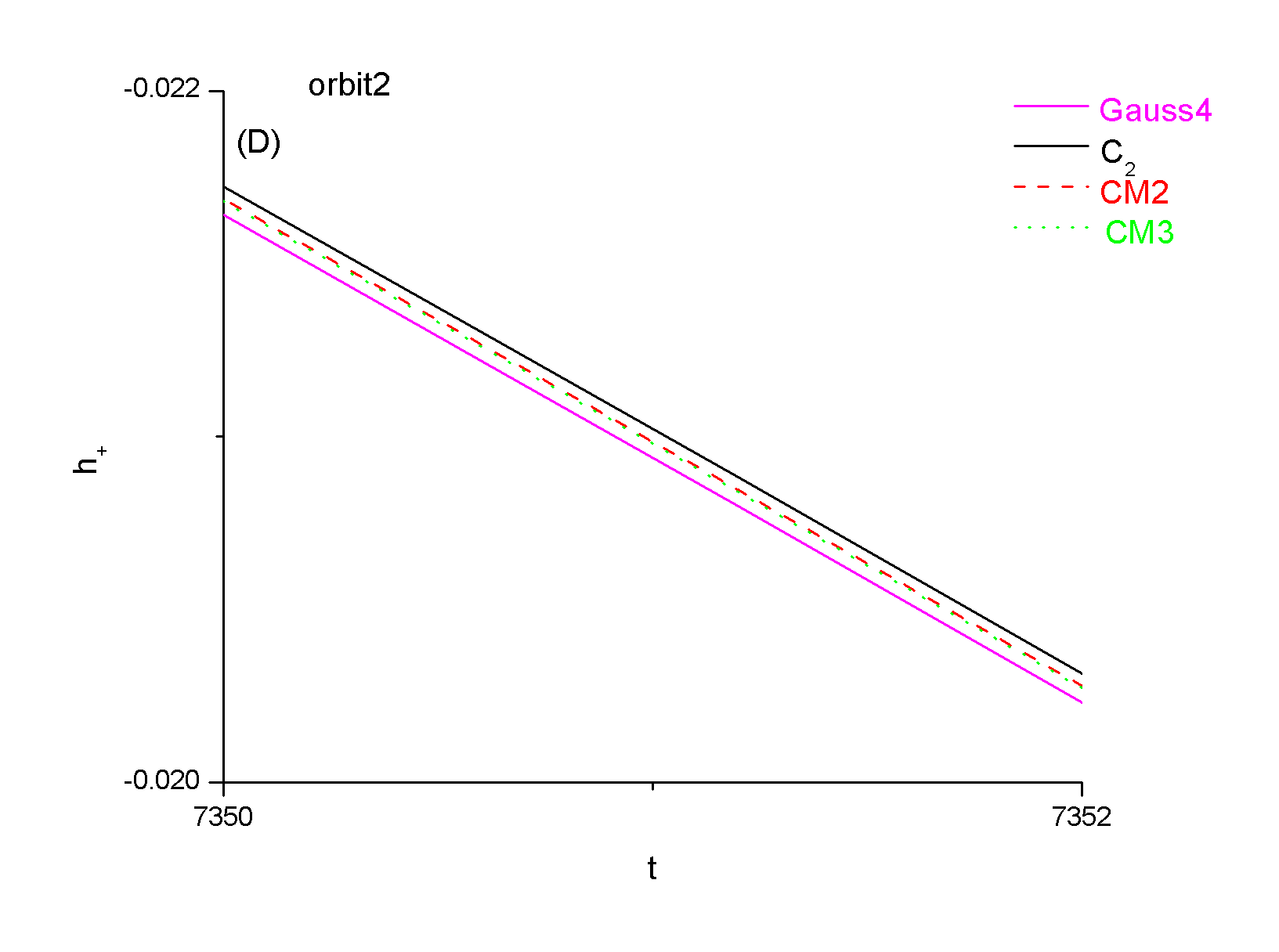}
\caption{The gravitational waveform $h_x$ and $h_+$ in orbit 2. (A). The global evolution diagram of $h_x$ drawn by $C_2$, $CM2$, $CM3$ and $Gauss4$, they are almost the same. (B). $h_x$ local magnification diagram, It can be seen that the waveforms of $CM3$ and $CM2$ are between that of $Gauss4$ and $C_2$. And $CM3$ is slightly closer to $Gauss4$ than $CM2$. (C). The global evolution graph of $h_ +$, with almost no difference among all algorithms. (D). $h_+$ locally enlarged figure shows that the rankings of evolution closest to $Gauss4$ are similar to the ranking in $h_x$.}
\label{fig9}
\end{figure}

Figure \ref{fig9} presents four distinct diagrams, labeled (A), (B), (C), and (D), each offering specific insights into the gravitational wave signatures generated by the studied algorithms with $\widehat{\mathbf{p}}=(1, 0, 0)$ and $\widehat{\mathbf{N}}=(0, sin(\pi /4), cos(\pi /4))$:

(A) and (C) show the overall gravitational waveforms for the '$h_x$' and '$h_+$' polarizations, respectively, providing a global perspective on the radiation emitted by the spinning compact binary during orbit 2. Across these comprehensive views, all algorithms yield waveforms that are essentially indistinguishable from one another, indicating a high degree of agreement in their overall representation of the gravitational wave signal.

(B) and (D) investigate the local amplification details of the '$h_x$' and '$h_+$' polarizations, where subtle differences in the waveforms become apparent. In these magnified sections, $CM3$ is revealed to exhibit the closest resemblance to the benchmark $Gauss4$ calculation, and slightly closer than $CM2$, with $C_2$ demonstrating the greatest deviation. This finding aligns with the conclusions drawn from the analysis of orbit 1.

\begin{figure}
\centering
\includegraphics[width=0.45\textwidth]{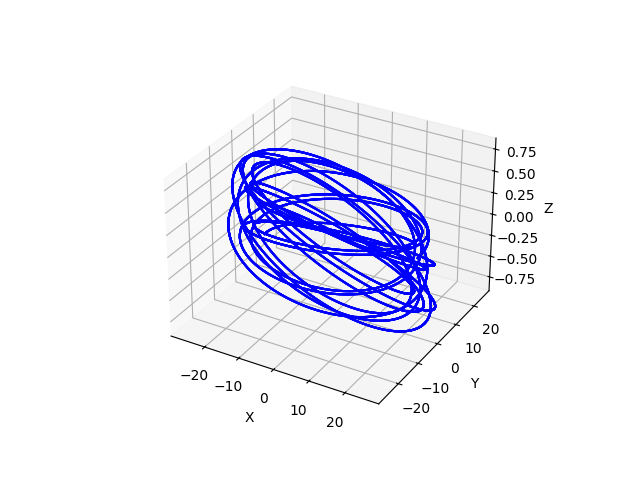}
\caption{The projection of orbit 2 onto $x-y-z$ space plot by $Gauss4$.}
\label{fig10}
\end{figure}

Lastly, Figure \ref{fig10} illustrates the orbital evolution of the binary system in configuration space as modeled by $Gauss4$ throughout orbit 2. This visualization offers a complementary perspective on the underlying dynamical processes driving the observed gravitational wave patterns, providing a more holistic understanding of the system's behavior.

Despite these distinctions, the general trends observed across the two orbits provide valuable insights into the strengths and weaknesses of the employed numerical methods, ultimately informing their suitability for accurately simulating a wide range of spinning compact binary configurations.

\section{Summary}\label{sec:4}
The present work has focused on the development and evaluation of the dissipated correction map method with the trapezoidal rule for numerical simulations of gravitational waves emitted by spinning compact binary systems. Our objective was to advance the frontiers of simulating inherently intricate and complex celestial phenomena, particularly concentrating on enhancing the precision and stability of extended phase-space algorithms. The proposed dissipated correction map builds upon the foundation of extended phase-space techniques, tackling key challenges encountered in modeling the complex dynamics of these systems using post-Newtonian approximations.  Here, we highlight the substantial enhancements offered by the dissipated correction map over the midpoint map and their implications for future research in gravitational wave astronomy.

In the numerical simulations and performance assessment,
we rigorously test the effectiveness of the $CM3$, we carried out extensive numerical simulations using a ten-dimensional canonical spin Hamiltonian, known for its tendency to display chaotic dynamics due to the absence of a fifth integral.  This choice of this Hamiltonian allowed us to appraise the correction map method's performance under conditions that are particularly taxing for numerical integrators.  The simulations were performed for Orbit 1 and 2, and the outcomes were contrasted against those obtained using the midpoint map which was the already established algorithms. Our analyses revealed several compelling advantages of the $CM3$ over the $C_2$ and alternative approaches.  Energy error evaluations demonstrated that the $CM3$ consistently achieved the lowest energy errors, signifying superior conservation attributes and a heightened level of accuracy in tracing the system's energy evolution.  Furthermore, the dissipated energy comparison showed that the $CM3$ closely paralleled the results obtained from the high-precision Gaussian algorithm, further supporting its accuracy in representing the dynamics of the binary system.
Temporal stability evaluations, quantified via phase space distance analysis, unmistakably showed that the $CM3$ outperformed the $CM2$ and $C_2$ over time. This means that the numerical solution of $CM3$ is closer to the $Gauss4$ algorithm as the truth value.

Visual scrutiny of the simulated gravitational waveforms provided compelling testament to the $CM3$ superiority.  Gravitational waveforms computed using the $CM3$ closely mimicked those generated by the high-precision Gaussian algorithm, particularly in regions of local magnification, indicating that the $CM3$ captures the fine-grained details of the wave signal with better performance. Therefore, compared with the previous $C_2$ algorithm, $CM2$ or $CM3$, especially $CM3$, is recommended to simulate the gravitational waveform of spinning compact binaries .

\end{document}